\documentclass[a4paper]{amsart}

  \expandafter\ifx\csname leqslant\endcsname\relax\usepackage{mathrsfs,txfonts}\fi
\usepackage{latexsym,multirow,multicol}\allowdisplaybreaks[1]
\usepackage[lutzsyntax]{virginialake}\vlsmallbrackets
\usepackage{hyperref}

\theoremstyle{plain}
\newtheorem{theorem}{Theorem}[section]
\newtheorem{corollary}[theorem]{Corollary}
\newtheorem{lemma}[theorem]{Lemma}

\theoremstyle{definition}
\newtheorem{definition}[theorem]{Definition}
\newtheorem{example}[theorem]{Example}

\newtheorem{remark}[theorem]{Remark}
\newtheorem{conjecture}[theorem]{Conjecture}

\begin{document}

\title{On the Proof Complexity of Deep Inference}

\author{Paola Bruscoli}
\author{Alessio Guglielmi}
\thanks{This research was partially supported by EPSRC grant EP/E042805/1 \emph{Complexity and Non-determinism in Deep Inference}.}
\date{\today}
\address{University of Bath, Bath BA2 7AY, UK\\
\url{http://cs.bath.ac.uk/pb/} and \url{http://alessio.guglielmi.name/res}}
\thanks{\copyright\ ACM, 2009. This is the authors' version of the work. It is posted here by permission of ACM for your personal use. Not for redistribution. The definitive version was published in \emph{ACM Transactions on Computational Logic} 10 (2:14) 2009, pp.\ 1--34, \url{http://doi.acm.org/10.1145/1462179.1462186}.}

\begin{abstract}
We obtain two results about the proof complexity of deep inference: 1) deep-inference proof systems are as powerful as Frege ones, even when both are extended with the Tseitin extension rule or with the substitution rule; 2) there are analytic deep-inference proof systems that exhibit an exponential speedup over analytic Gentzen proof systems that they polynomially simulate.
\end{abstract}

\maketitle

\section{Introduction}

\newcommand{\SKS}{\mathsf{SKS}}
\emph{Deep inference} is a relatively new methodology in proof theory, consisting in dealing with proof systems whose inference rules are applicable at any depth inside formulae \cite{Gugl:06:A-System:kl}. We obtain two results about the proof complexity of deep inference:
\begin{itemize}
\item deep-inference proof systems are as powerful as Frege ones, even when both are extended with the Tseitin extension rule or with the substitution rule;
\item there are analytic deep-inference proof systems that exhibit an exponential speed\-up over analytic Gentzen proof systems that they polynomially simulate.
\end{itemize}
These results are established for the \emph{calculus of structures}, or \emph{CoS}, the simplest formalism in deep inference \cite{Gugl:06:A-System:kl}, and in particular for its proof system $\SKS$, introduced by Br\"unnler in \cite{Brun:04:Deep-Inf:rq} and then extensively studied \cite{Brun:03:Atomic-C:oz,Brun:03:Two-Rest:mn,Brun:06:Cut-Elim:cq,Brun:06:Locality:zh,BrunGugl:04:A-First-:ys,BrunTiu:01:A-Local-:mz}.

Our contributions fit in the following picture.
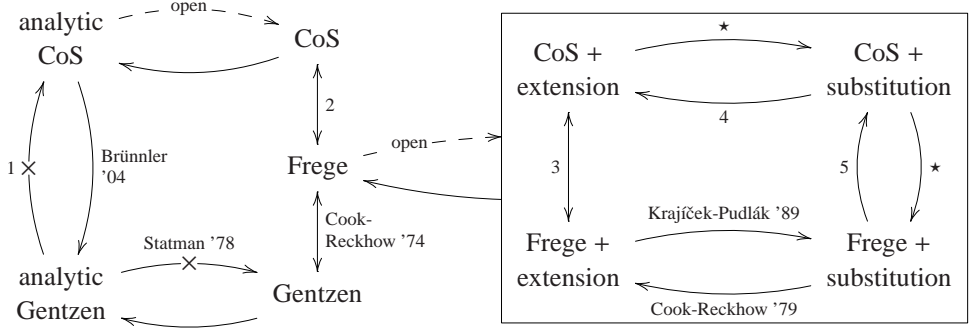
\[\hss
\def\diagboxaux #1{\hbox{\kern.5pc #1\strut\kern.5pc}}
\def\diagbox #1{\vbox{\kern.25pc\diagboxaux{#1}\kern.25pc}}
\def\diagboxx #1#2{\vbox{\kern.25pc\halign{\hfil##\hfil\cr\diagboxaux{#1}\cr\diagboxaux{#2}\cr}\kern.25pc}}
\xy<1pc,0pt>:<0pt,1pc>::
\POS(13,0)*{\xybox{<1pc,0pt>:<0pt,.75pc>::
	\POS(0,8)*{\diagboxx{CoS +}{extension}}="A"
	\POS(10,8)*{\diagboxx{CoS +}{substitution}}="B"
	\POS(0,0)*{\diagboxx{Frege +}{extension}}="C"
	\POS(10,0)*{\diagboxx{Frege +}{substitution}}="D"
	\POS"A"
	\ar@{->}@/^1pc/"B"^{\!\!\!\star}
	\POS"B"
	\ar@{->}@/^1pc/"A"^{\!\!\!4}
	\POS"A"
	\ar@{<->}"C"_{3}
	\POS"C"
	\ar@{->}@/^1pc/"D"^{\mbox{\scriptsize Kraj\'i\v cek-Pudl\'ak '89\ \ }}
	\POS"B"
	\ar@{->}@/^1pc/"D"^{\star}
	\POS"D"
	\ar@{->}@/^1pc/"B"^{5}
	\POS"D"
	\ar@{->}@/^1pc/"C"^{\mbox{\scriptsize Cook-Reckhow '79\ \ }}}
	}="E"\POS"E"*\frm{-}
\POS(0,0)*{\diagbox{Frege}}="F"
\POS(0,4)*{\diagbox{CoS}}="G"
\POS(0,-4)*{\diagbox{Gentzen}}="H"
\POS"F"
\ar@{<-}@/_1pc/"E"
\ar@{-->}@/^1pc/"E"|<<<<{\mbox{\scriptsize\ open\ }}
\POS"F"
\ar@{<->}"G"_{2}
\POS"F"
\ar@{<->}"H"^{\vbox{\offinterlineskip\hbox{\scriptsize Cook-\strut}
                    \hbox{\scriptsize Reckhow '74\strut}}}
\POS(-8,4)*{\diagboxx{analytic}{CoS}}="I"
\POS(-8,-4)*{\diagboxx{analytic}{Gentzen}}="J"
\POS"I"
\ar@{->}@/^1pc/"J"^{\vbox{\offinterlineskip\hbox{\scriptsize Br\"unnler\strut}
                    \hbox{\scriptsize '04\strut}}}
\ar@{<-}@/_1pc/"J"_{1\;}|{\mbox{$\times$}}
\POS"J"
\ar@{->}@/^1pc/"H"^{\mbox{\scriptsize Statman '78\strut}}|{\mbox{$\times$}}
\ar@{<-}@/_1pc/"H"
\POS"I"
\ar@{-->}@/^1pc/"G"|{\mbox{\scriptsize\ open\ }}
\ar@{<-}@/_1pc/"G"
\endxy\;\;
\hss\]
The notation $\xy<1pc,0pt>:<0pt,1pc>::(0,0)*{\mathscr{F}\strut\;}\ar@{->}(2.5,0)*{\;\mathscr{G}\strut}\endxy$ indicates that formalism $\mathscr{F}$ polynomially simulates formalism $\mathscr{G}$; the notation $\xy<1pc,0pt>:<0pt,1pc>::(0,0)*{\mathscr{F}\strut\;}\ar@{->}|<<<{\mbox{$\times$}}(2.5,0)*{\;\mathscr{G}\strut}\endxy$ indicates that it is known that this does not happen.

The left side of the picture represents, in part, the following. Analytic Gentzen systems, \emph{i.e.}, Gentzen proof systems without the cut rule, can only prove certain formulae, which we call `Statman tautologies', with proofs that grow exponentially in the size of the formulae. On the contrary, Gentzen systems with the cut rule can prove Statman tautologies by polynomially growing proofs. So, Gentzen systems p-simulate analytic Gentzen ones, but not vice versa \cite{Stat:78:Bounds-f:fj}. Cook and Reckhow proved that Frege and Gentzen systems are polynomially equivalent, \emph{i.e.}, each Frege system polynomially simulates any Gentzen system and vice versa \cite{CookReck:74:On-the-L:lr}.

In the box at the right of the figure, `extension' refers to the Tseitin extension rule, and `substitution' to the substitution rule. The works of Cook and Reckhow \cite{CookReck:79:The-Rela:mf} and Kraj\'i\v cek and Pudl\'ak \cite{KrajPudl:89:Proposit:fk} established that Frege + extension and Frege + substitution are polynomially equivalent. It is immediate to see that these formalisms polynomially simulate Frege and Gentzen, which, in turn, polynomially simulate analytic Gentzen. It is a major open problem to establish whether Frege polynomially simulates Frege + extension/substitution.

In this work, we establish the following results (numbered as in the previous figure):
\begin{enumerate}
\item Analytic Gentzen does not polynomially simulate analytic CoS (essentially in the form of system $\SKS$ without cut); in fact, Statman tautologies admit polynomially growing proofs in analytic CoS (Theorem~\ref{ThStat}).
\item CoS and Frege are polynomially equivalent (Theorems~\ref{CorSKSPSimFR} and \ref{CorFRPSimSKS}).
\item There is a natural notion of (Tseitin) extension for CoS, and CoS + extension is polynomially equivalent to Frege + extension (Theorem~\ref{TheoExtFRPEqSKSX}).
\item There is a natural notion of substitution for CoS, and CoS + substitution polynomially simulates CoS + extension (Theorem~\ref{CorSKSSubPSimExtSKS}).
\item Frege + substitution polynomially simulates CoS + substitution; this way, we know that all extended formalisms are polynomially equivalent (Theorem \ref{CorFRSubPSimSKSSub}).
\end{enumerate}
The polynomial simulations indicated by $\star$ arcs in the picture follow from the others.

Establishing whether analytic CoS polynomially simulates CoS is an open problem.

After the necessary preliminaries, in Section~\ref{SectPrel}, we see how CoS expresses Gentzen systems, including their properties, like analyticity, and then in Section~\ref{SectStatman} how it provides for exponentially more compact proofs. The relation between CoS and Frege systems is explored in Section~\ref{SectFrege} and the extensions are studied in Section~\ref{SectExt}. We conclude the article with a list of open problems, in Section~\ref{SectOpen}.

\section{Preliminaries}\label{SectPrel}

In this section, we quickly introduce the necessary deep-inference notions. A more extensive treatment of much of this material is in Br\"unnler's \cite{Brun:04:Deep-Inf:rq}.

We only need the following, basic proof complexity notions (see \cite{CookReck:79:The-Rela:mf}).

\begin{definition}
A (\emph{propositional}) \emph{proof system} is a binary relation $\mathcal S$ between \emph{formulae} $\alpha$ and \emph{proofs} $\Pi$ such that $\mathcal S$ is computable in polynomial time, and the formula $\alpha$ is a tautology if and only if there is a proof $\Pi$ such that ${\mathcal S}(\alpha,\Pi)$; in this case we say that $\Pi$ is a proof of $\alpha$ in $\mathcal S$. We say that proof system $\mathcal S$ \emph{p-simulates} proof system $\mathcal S'$ if there is a polynomial-time computable algorithm that transforms every proof in $\mathcal S'$ into a proof in $\mathcal S$ of the same tautology. Two proof systems are \emph{p-equivalent} if each p-simulates the other.
\end{definition}

\begin{remark}
In the following, we state theorems on the existence of proofs in one proof system when proofs exist in another proof system, such that their size is polynomially related. Implicitly, we always mean that the new proofs are obtained by transforming the old ones by way of a polynomial-time computable algorithm.
\end{remark}

Deep inference is a relatively recent development in proof theory. Its main idea is to provide a finer analysis of inference than possible with traditional methods, and one of the main objectives is to obtain a geometric semantics for proofs, inspired by linear logic's proof nets \cite{Gira:87:Linear-L:wm}. Another objective is to provide a uniform and useful syntactic treatment of several logics, especially modal ones, for which no satisfactory proof theory existed before.

In deep inference, several formalisms can be defined with excellent structural properties, like locality for all the inference rules. The calculus of structures \cite{Gugl:06:A-System:kl} is one of them and is now well developed for classical \cite{Brun:03:Atomic-C:oz,Brun:06:Cut-Elim:cq,Brun:06:Locality:zh,BrunTiu:01:A-Local-:mz}, intuitionistic \cite{Tiu:06:A-Local-:gf}, linear \cite{Stra:02:A-Local-:ul,Stra:03:MELL-in-:oy}, modal \cite{Brun::Deep-Seq:ay,GoreTiu:06:Classica:uq,Stou:06:A-Deep-I:rt} and commutative/noncommutative logics \cite{Gugl:06:A-System:kl,Tiu:06:A-System:ai,Stra:03:Linear-L:lp,Brus:02:A-Purely:wd,Di-G:04:Structur:wy,GuglStra:01:Non-comm:rp,GuglStra:02:A-Non-co:lq,GuglStra:02:A-Non-co:dq,Kahr:06:Reducing:hc,Kahr:07:System-B:fk}; for all these logics, quantification can be defined at any order. We emphasise that deep inference is developing the first reasonable proof theory for modal logics; the large number of different modal logic systems can be studied in simple and modular deep-inference systems, which are similar to their propositional logic counterparts and enjoy the same locality properties. The calculus of structures promoted the discovery of a new class of proof nets for classical and linear logic \cite{LamaStra:05:Construc:qq,LamaStra:05:Naming-P:ov,LamaStra:06:From-Pro:et,StraLama:04:On-Proof:ec} (see also \cite{Guir:06:The-Thre:qt}). Moreover, there exist implementations in Maude of deep-inference proof systems \cite{Kahr:07:Maude-as:lr}.

In this article, we focus on the calculus of structures because it is well developed and is probably the simplest formalism definable in deep inference. The complexity results that we present here are not dependent on the choice of formalism; rather, they only depend on the deep-inference methodology and the finer granularity of inference rules that it yields. Adopting deep inference basically means that it is possible to replace subformulae inside formulae by other, implied subformulae, and that there is no limit to the nesting depth of subformulae. Formalisms like Gentzen's sequent calculus differ because they only rewrite formulae, or sequents, around their root connectives, and (we argue) they suffer excessive rigidity in the syntax and they do not sufficiently support geometric semantics.

Because of its geometric nature, it is important, in deep inference, to control whether propositional variables can be instantiated by formulae. In particular, normalisation (cut elimination) in deep inference crucially depends on the availability of `atomic' inference rules, which are rules related to some topological invariants (see, for example, \cite{GuglGund:07:Normalis:lr} for normalisation in propositional logic). In practice, we need two kinds of propositional variables: the atoms, only subject to renaming, and the formula variables, subject to (unrestricted) substitution. This distinction does not bear dramatic effects on proof complexity, but it does allow for some finer measures than otherwise possible.

\newcommand{\fff }{\mathsf f}
\newcommand{\ttt }{\mathsf t}
\newcommand{\size}[1]{{\vert}#1{\vert}}
\begin{definition}
\emph{Formulae} of the \emph{calculus of structures}, or \emph{CoS}, are denoted by $\alpha$, $\beta$, $\gamma$, $\delta$ and are freely built from: \emph{units}, like $\fff$ (false) and $\ttt$ (true); \emph{atoms} $a$, $b$, $c$, $d$ and $\bar a$, $\bar b$, $\bar c$, $\bar d$; (\emph{formula}) \emph{variables} $A$, $B$, $C$, $D$ and $\bar A$, $\bar B$, $\bar C$, $\bar D$; \emph{logical relations}, like \emph{disjunction} $\vlsbr[\alpha.\beta]$ and \emph{conjunction} $\vlsbr(\alpha.\beta)$. A formula is \emph{ground} if it contains no variables. We usually omit external brackets of formulae, and sometimes we omit dispensable brackets under associativity. We use $\equiv$ to denote literal equality of formulae. The \emph{size} $\size\alpha$ of a formula $\alpha$ is the number of unit, atom, and variable occurrences appearing in it. On the set of atoms, there is an involution $\bar\cdot$, called \emph{negation} (\emph{i.e.}, $\bar\cdot$ is a bijection from the set of atoms to itself such that $\bar{\bar a}\equiv a$); we require that $\bar a\not\equiv a$ for every $a$; when both $a$ and $\bar a$ appear in a formula, we mean that atom $a$ is mapped to $\bar a$ by $\bar\cdot$. An analogous involution is defined on the set of formula variables. The (\emph{De Morgan}) \emph{dual} of a formula is obtained by exchanging disjunction and conjunction and applying negation to all atoms and variables; we denote duals by using $\bar\cdot$; for example, the De Morgan dual of $\alpha\equiv\vls[\ttt.(a.[\bar B.c])]$ is $\bar\alpha\equiv\vls(\fff.[\bar a.(B.\bar c)])$. A \emph{context} is a formula where one \emph{hole} $\vlhole$ appears in the place of a subformula; for example, $\vls[A.(b.\vlhole)]$ is a context; the generic context is denoted by $\xi\vlhole$. The hole can be filled with formulae; for example, if $\xi\vlhole\equiv\vls(b.[\vlhole.c])$, then $\xi\{a\}\equiv\vls(b.[a.c])$, $\xi\{b\}\equiv\vls(b.[b.c])$ and $\xi\{\vls(a.B)\}\equiv\vls(b.[(a.B).c])$. The \emph{size} of $\xi\vlhole$ is defined as $\size{\xi\vlhole}=\size{\xi\{a\}}-1$.
\end{definition}

\begin{remark}
We do not say that $a$ is positive and $\bar a$ is negative. It only matters that, when $a$ and $\bar a$ appear in the same formula, if one is negative the other is positive. In absence of disambiguating information, there are two ways in which $\xi\{b\}$ might correspond to $\vls(b.[b.c])$: one such that $\xi\{a\}\equiv\vls(a.[b.c])$ and another such that $\xi\{a\}\equiv\vls(b.[a.c])$.
\end{remark}

The language of formulae is redundant because we can choose whether to use atoms or formula variables whenever a propositional variable is needed. The distinction between atoms and formula variables only plays a role in the choice of applicable inference rules, and this aspect is controlled by renamings and substitutions.

\newcommand{\ot}{\mathord/}
\begin{definition}
A \emph{renaming} is a map from the set of atoms to itself, and is denoted by $\{a_1\ot b_1,a_2\ot b_2,\dots\}$; we use $\rho$ for renamings. A renaming of $\alpha$ by $\rho=\{a_1\ot b_1,a_2\ot b_2,\dots\}$ is indicated by $\alpha\rho$ and is obtained by simultaneously substituting every occurrence of $a_i$ in $\alpha$ by $b_i$ and every occurrence of $\bar a_i$ by $\bar b_i$; for example, if $\alpha\equiv\vls(a.[b.(a.[\bar a.C])])$ then $\alpha\{a\ot\bar b,\bar b\ot c\}\equiv\vls(\bar b.[\bar c.(\bar b.[b.C])])$. A \emph{substitution} is a map from the set of formula variables to formulae, denoted by $\{A_1\ot\beta_1,A_2\ot\beta_2,\dots\}$; we use $\sigma$ for substitutions. An \emph{instance} of $\alpha$ by $\sigma=\{A_1\ot\beta_1,A_2\ot\beta_2,\dots\}$ is indicated by $\alpha\sigma$ and is obtained by simultaneously substituting every occurrence of variable $A_i$ in $\alpha$ by formula $\beta_i$ and every occurrence of $\bar A_i$ by the De Morgan dual of $\beta_i$; for example, if $\alpha\equiv\vls[(A.[A.\bar A]).b]$ then $\alpha\{A\ot\vlsbr(c.\bar B)\}\equiv\vls[((c.\bar B).[(c.\bar B).[\bar c.B]]).b]$.
\end{definition}

\newcommand  {\gi }{\mathsf{i}}
\newcommand  {\gid}{{\gi{\downarrow}}}
\newcommand  {\ai }{\mathsf{ai}}
\newcommand  {\aid}{{\ai{\downarrow}}}
\newcommand  {\gc }{\mathsf{c}}
\renewcommand{\gcd}{{\gc{\downarrow}}}
\begin{definition}
A CoS (\emph{inference}) \emph{rule} $\nu$ is an expression $\vlsmash{\vlinf{\nu}{}\beta\alpha}$, where formulae $\alpha$ and $\beta$ are called \emph{premiss} and \emph{conclusion}, respectively. A (\emph{rule}) \emph{instance} $\vlinf{\nu}{}\delta\gamma$ of $\vlinf{\nu}{}\beta\alpha$ is such that $\gamma\equiv\alpha\rho\sigma$ and $\delta\equiv\beta\rho\sigma$, for some renaming $\rho$ and substitution $\sigma$. For some context $\xi\vlhole$, a CoS (\emph{inference}) \emph{step}, \emph{generated by} rule $\vldownsmash{\vlinf{\nu}{}\beta\alpha}$ \emph{via} its instance $\vldownsmash{\vlinf{\nu}{}\delta\gamma}$, is the expression $\vlinf{\nu}{}{\xi\{\delta\}}{\xi\{\gamma\}}$.
\end{definition}

\begin{example}
Given $\vlsmash{\vlinf{\gid}{}{\vls[A.\bar A]}{\ttt}}$ and $\vlsmash{\vlinf{\aid}{}{\vls[a.\bar a]}{\ttt}}$, then $\vlinf{\gid}{}{\vls[(A.b).[\bar A.\bar b]]}{\ttt}$ is an instance of $\gid$ and $\vlinf\nu{}{\vls[b.\bar b]}{\ttt}$ is an instance of both $\gid$ and $\aid$. The rule $\vlinf{\gcd}{}A{\vls[A.A]}$ generates the inference step $\vlinf{\gcd}{}{\vls([a.b].(c.D))}{\vls([a.b].[(c.D).(c.D)])}$.
\end{example}

We usually classify deep-inference rules in three classes. For this, we rely on the notion of linearity, which, in this context, essentially means the same as in term-rewriting: a rewriting rule is linear if variables appear once in both sides of the rule. In other words, a linear rule does not create or destroy anything. These are the three classes of rules:
\begin{enumerate}
\item\emph{Atomic rules}. They usually correspond to structural rules in Gentzen systems; in normalisation and in semantics of proofs, they play a crucial role because they express the causality relations between atoms, so shaping the geometry of proofs. Their instances are obtained by renaming.
\item\emph{Noninvertible linear rules}. They usually correspond to logical rules in Gentzen systems. Since they are noninvertible, they express proper inference choices, but since they are linear, they do not alter the geometry of causality between atoms. Their instances are obtained by substitution.
\item\emph{Invertible linear rules}. These rules are equivalences between formulae that do not correspond to proper inference choices and have no impact on the geometry of proofs. For this reason, they are usually gathered into one big equivalence relation between formulae, corresponding to just one rule, defined via substitution.
\end{enumerate}
The success of deep inference is due to its ability to separate rules into classes~1 and 2, which is only possible by adopting deep inference. The references to the `geometry of proofs' can be understood by reading \cite{GuglGund:07:Normalis:lr,LamaStra:05:Naming-P:ov}. Class~3 allows us to greatly simplify proofs and to hide, so to speak, a great deal of logical complexity (in the sense of size of proofs). We start by defining our `class~3' rule, the others being dependent on specific proof systems.

\begin{definition}
The equality relation $=$ on formulae is defined by closing the equations in Figure~\ref{FigEq} by reflexivity, symmetry, transitivity and by applying context closure. We define the inference rule $=$ as $\vlinf{=}{}\beta\alpha$, where $\alpha=\beta$.
\end{definition}

\begin{figure}[t]
\[
\begin{array}{cc}
\begin{array}{c}
\mbox{\emph{Commutativity}}\\
\noalign{\bigskip}
\vls[\alpha.\beta]=[\beta.\alpha]\\
\vls(\alpha.\beta)=(\beta.\alpha)\\
\noalign{\bigskip}
\mbox{\emph{Associativity}}\\
\noalign{\bigskip}
\vls[[\alpha.\beta].\gamma]=[\alpha.[\beta.\gamma]]\\
\vls((\alpha.\beta).\gamma)=(\alpha.(\beta.\gamma))\\
\end{array}
&\qquad
\begin{array}{c}
\mbox{\emph{Units}}\\
\noalign{\bigskip}
\vls[\alpha.\fff]=\alpha\\
\vls(\alpha.\ttt)=\alpha\\
\vls[\ttt.\ttt]=\ttt\\
\vls(\fff.\fff)=\fff\\
\noalign{\bigskip}
\mbox{\emph{Context closure}}\\
\noalign{\bigskip}
\mbox{if $\alpha=\beta$ then $\xi\{\alpha\}=\xi\{\beta\}$}\\
\end{array}
\end{array}
\]
\caption{Equality $=$ on formulae.}
\label{FigEq}
\end{figure}

The following remark helps in assessing how much complexity is hidden in $=$.

\newcommand{\Ord}[1]{{\mathsf O}(#1)}
\begin{remark}\label{RemEq}
It is possible to decide $\alpha=\beta$ in polynomial time by reducing $\alpha$ and $\beta$ to some canonical form and comparing the canonical forms. A canonical form under $=$ of any given formula can be obtained, for example, by removing as many units as possible and ordering units, atoms, and variables according to an arbitrary order; the canonical form is normal for associativity and units equations, when these are orientated from left to right. Let us assume a total order on the set of units, atoms, and variables; we now see in detail how to find an equivalent canonical formula in the case of a formula only containing one logical relation. On the formula, use commutativity until the minimal unit, atom, or variable appears in the leftmost position, then use associativity, orientated from left to right, until normality is reached. For example, on $\vls[[b.d].[c.a]]$, we perform the steps
\[
\vls[[b.d].[c.a]]\leadsto[[c.a].[b.d]]
                 \leadsto[[a.c].[b.d]]
                 \leadsto[a.[c.[b.d]]]\quad.
\]
This phase requires $\Ord{n}$ steps, where $n$ is the size of the formula. We proceed the same way on the subformula immediately following the first element, and so on recursively; for example,
\[
\vls[a.[c.[b.d]]]\leadsto[a.[[b.d].c]]
                 \leadsto[a.[b.[d.c]]]
                 \leadsto[a.[b.[c.d]]]\quad.
\]
The number of steps of the algorithm for a formula only containing one logical relation is then $\Ord{n^2}$. On a generic formula, the same algorithm can be used, with the same number-of-steps complexity $\Ord{n^2}$ on the size $n$ of the given formula, by adopting the lexicographic order induced by the given total order. This is an example, also involving an initial $\Ord{n}$ phase of simplification of units:
\vlstore{
       &\vls((\ttt.\ttt).[[a.([[b.d].[c.a]].[[B.c].[\ttt.a]])].\fff]     )\\
{}\leadsto{}&\vls( \ttt .[[a.([[b.d].[c.a]].[[B.c].[\ttt.a]])].\fff]     )\\
{}\leadsto{}&\vls(       [[a.([[b.d].[c.a]].[[B.c].[\ttt.a]])].\fff].\ttt)\\
{}\leadsto{}&\vls        [[a.([[b.d].[c.a]].[[B.c].[\ttt.a]])].\fff]      \\
{}\leadsto{}&\vls         [a.([[b.d].[c.a]].[[B.c].[\ttt.a]])]            \\
{}\leadsto^\star{}
       &\vls              [a.([a.[b.[c.d]]].[[B.c].[\ttt.a]])]            \\
{}\leadsto^\star{}
       &\vls              [a.([a.[b.[c.d]]].[\ttt.[a.[c.B]]])]            \\
{}\leadsto{}&\vls         [a.([\ttt.[a.[c.B]]].[a.[b.[c.d]]])]            \\
{}\leadsto{}&\vls         [  ([\ttt.[a.[c.B]]].[a.[b.[c.d]]]).a]
\quad\vlnos.}
\begin{align*}
\vlread
\end{align*}
This way we obtain a (unique, of course) canonical formula in $\Ord{n^2}$ steps, given any formula of size $n$, so we can decide the equivalence of two formulae $\alpha$ and $\beta$ in $\Ord{n^2}$ steps, where $n=\size\alpha+\size\beta$. Notice that at each step the size of the formula stays the same or diminishes.
\end{remark}

\begin{definition}
A CoS (\emph{proof}) \emph{system} is a finite set of inference rules. A CoS \emph{derivation} $\Phi$ of \emph{length} $k$ in proof system $\mathcal S$, whose \emph{premiss} is $\alpha_0$ and \emph{conclusion} is $\alpha_k$, is a chain of inference steps
\[
\Phi=
\vlderivation{
\vlin{\nu_k    }{}{\alpha_k    }   {
\vlin{\nu_{k-1}}{}{\alpha_{k-1}}  {
\vlin{\nu_2    }{}{\vdots      } {
\vlin{\nu_1    }{}{\alpha_1    }{
\vlhy             {\alpha_0    }}}}}}
\quad,
\]
such that $\nu_1,\dots,\nu_k$ is a sequence of inference rules that alternate between the $=$ rule and any rule of system $\mathcal S$, where $k\ge0$. The same derivation can be indicated by $\vlder{\Phi}{\mathcal S}{\alpha_k}{\alpha_0}$, when the details are known or irrelevant; a \emph{proof} is a derivation whose premiss is $\ttt$. A derivation is \emph{ground} if it contains no variables. Sometimes, we omit to indicate the inference steps generated by $=$. The \emph{size} $\size\Phi$ of derivation $\Phi$ is the number of unit, atom, and variable occurrences appearing in it. We denote by $\xi\{\Phi\}$ the result of including every formula of $\Phi$ into the context $\xi\vlhole$. We denote by $\Phi\rho$ and $\Phi\sigma$ the expression obtained from $\Phi$ by applying renaming $\rho$ and substitution $\sigma$ to every formula in $\Phi$. A CoS proof system that, for every valid implication $\vls\alpha\vlim\beta$, contains a derivation with premiss $\alpha$ and conclusion $\beta$, is said to be \emph{implicationally complete}.
\end{definition}

\begin{remark}\label{RemClosure}
If $\Phi$ is a derivation, then $\xi\{\Phi\}$, $\Phi\rho$, and $\Phi\sigma$ are derivations, for every context $\xi\vlhole$, renaming $\rho$, and substitution $\sigma$.
\end{remark}

We use the notion of groundness to relate the complexity of deep-inference proof systems with atomic rules to proof systems without atomic rules, including those outside of deep inference. Due to the aforementioned redundancy in the language, groundness is not really a restriction.

\begin{remark}\label{RemGround}
Every nonground derivation can be transformed into an equivalent,\break ground one, by replacing variables with atoms in such a way that newly introduced atoms are different from the already present one.
\end{remark}

\newcommand{\SKSg}{\mathsf{SKSg}}
We can now define some deep-inference proof systems. System $\SKS$ is the most important for the proof theory of classical logic, because of its atomic structural rules. System $\SKSg$ relates $\SKS$ to proof systems in other formalisms, like Frege.

\newcommand{\gw  }{\mathsf{w}}
\newcommand{\gwd }{{\gw{\downarrow}}}
\newcommand{\giu }{{\gi{\uparrow}}}
\newcommand{\gwu }{{\gw{\uparrow}}}
\newcommand{\gcu }{{\gc{\uparrow}}}
\newcommand{\swi }{\mathsf{s}}
\newcommand{\KSg }{\mathsf{KSg}}
\newcommand{\aw }{\mathsf{aw}}
\newcommand{\ac }{\mathsf{ac}}
\newcommand{\awd}{{\aw{\downarrow}}}
\newcommand{\acd}{{\ac{\downarrow}}}
\newcommand{\aiu}{{\ai{\uparrow}}}
\newcommand{\awu}{{\aw{\uparrow}}}
\newcommand{\acu}{{\ac{\uparrow}}}
\newcommand{\med}{\mathsf{m}}
\newcommand{\KS }{\mathsf{KS}}
\begin{definition}\label{DefSKS}
CoS proof systems $\KSg=\{\gid,\gwd,\gcd,\swi\}$, $\SKSg=\KSg\cup\{\giu,\gwu,\linebreak\gcu\}$, $\KS=\{\aid,\awd,\acd,\swi,\med\}$ and $\SKS=\KS\cup\{\aiu,\awu,\acu\}$ are defined in Figures~\ref{FigSKSg} and \ref{FigSKS}, for a language containing $\fff$, $\ttt$, disjunction, and conjunction. Proof systems where none of the rules $\giu$, $\aiu$, $\gwu$, and $\awu$ appear are said to be \emph{analytic}.
\end{definition}

\begin{figure}[t]
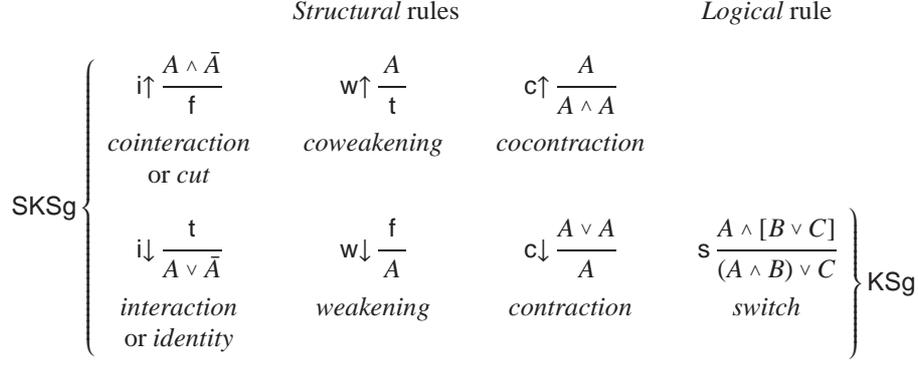

\[\hss
\begin{array}{cccccc}
   &\multispan3{\hfil\emph{Structural} rules\hfil}%
      &\quad\mbox{\emph{Logical} rule}                                        \\
\noalign{\bigskip}
\multirow{10}{*}[9.5pt]{$\SKSg\left\{\vbox to60pt{\vss}\right.\!\!\!\!$}
   &           \vlinf\giu{}\fff{\vls(A.{\bar A})}
      &   \quad\vlinf\gwu{}\ttt A
         &\quad\vlinf\gcu{}{\vls(A.A)}A                                       \\
\noalign{\smallskip}
   &\emph{cointeraction}
      &   \quad\emph{coweakening}
         &\quad\emph{cocontraction}                                           \\
   &\mbox{or \emph{cut}}                                                      \\
\noalign{\bigskip}
   &              \vlinf\gid{}{\vls[A.{\bar A}]}\ttt
      &      \quad\vlinf\gwd{}A\fff
         &   \quad\vlinf\gcd{}A{\vls[A.A]}
            &\quad\vlinf\swi{}{\vls[(A.B).C]}
                               {\vls(A.{[B.C]})}
               &\multirow{5}{*}[12.5pt]{$\!\!\!\!\left.\vbox 
                                                    to32pt{\vss}\right\}\KSg$}\\
\noalign{\smallskip}
   &              \emph{interaction}
      &      \quad\emph{weakening}
         &   \quad\emph{contraction}
            &\quad\emph{switch}                                               \\
   &\mbox{or \emph{identity}}                                                 \\
\end{array}
\hss\]
\caption{Systems $\SKSg$ and $\KSg$.}
\label{FigSKSg}
\end{figure}

\begin{figure}[t]
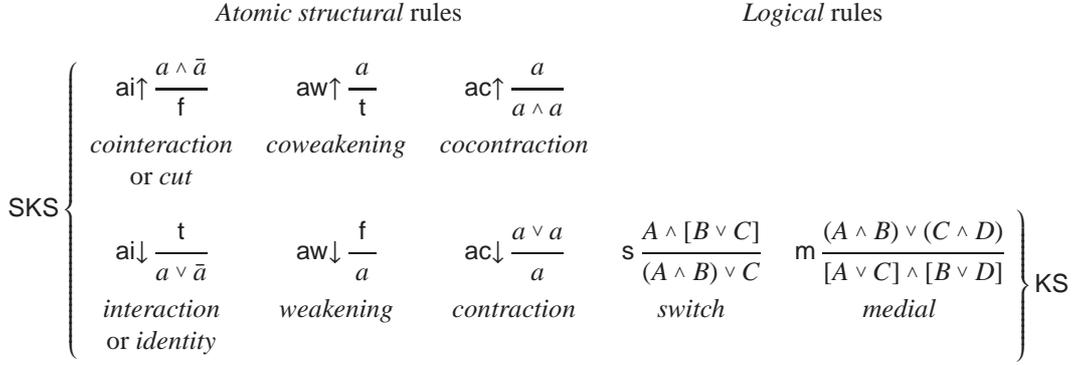

\[\hss
\begin{array}{ccccccc}
   &   \multispan3{\hfil  \emph{Atomic structural} rules\hfil}%
      &\multispan2{\hfil\;\emph{Logical} rules\hfil                          }\\
\noalign{\bigskip}
\multirow{10}{*}[9.5pt]{$\SKS\left\{\vbox to60pt{\vss}\right.\!\!\!\!$}
   &        \vlinf\aiu{}\fff{\vls(a.{\bar a})}
      &   \;\vlinf\awu{}\ttt a
         &\;\vlinf\acu{}{\vls(a.a)}a                                          \\
\noalign{\smallskip}
   &\emph{cointeraction}
      &   \;\emph{coweakening}
         &\;\emph{cocontraction}                                              \\
   &\mbox{or \emph{cut}}                                                      \\
\noalign{\bigskip}
   &              \vlinf\aid{}{\vls[a.{\bar a}]}\ttt
      &         \;\vlinf\awd{}a\fff
         &      \;\vlinf\acd{}a{\vls[a.a]}
            &   \;\vlinf\swi{}{\vls[(A.B).C]}
                                 {\vls(A.{[B.C]})}
               &\;\vlinf\med{}{\vls([A.C].{[B.D]})}
                                 {\vls[(A.B).{(C.D)}]}
                  &\multirow{5}{*}[12.5pt]{$\!\!\!\!\left.\vbox 
                                                     to32pt{\vss}\right\}\KS$}\\
\noalign{\smallskip}
   &              \emph{interaction}
      &         \;\emph{weakening}
         &      \;\emph{contraction}
            &   \;\emph{switch}
               &\;\emph{medial}                                               \\
   &\mbox{or \emph{identity}}                                                 \\
\end{array}
\hss\]
\caption{Systems $\SKS$ and $\KS$.}
\label{FigSKS}
\end{figure}

\begin{example}\label{ExCentrSt}
This is a valid derivation in all CoS proof systems defined previously (and it plays a role in the proof of Lemma~\ref{LemStat}):
\[
\vlderivation                                                               {
\vlin{=   }{}{\vls[[\bar\alpha.\gamma].[((\alpha.c).(\alpha.d)).\delta]]}  {
\vlin{\swi}{}{\vls[\gamma.[[(((\alpha.d).c).\alpha).\bar\alpha].\delta]]} {
\vlin{=   }{}{\vls[\gamma.[(((\alpha.d).c).[\alpha.\bar\alpha]).\delta]]}{
\vlhy        {\vls[\gamma.[(([\bar\alpha.\alpha].c).(\alpha.d)).\delta]]}}}}}
\quad.
\]
\end{example}

Note that $\SKSg$, $\KSg$, $\SKS$, and $\KS$ are closed under renaming and substitution (see Remark~\ref{RemClosure}). This is so because of the distinction between atoms and formula variables. Obtaining the closure of these and other systems under renaming and substitution is one of the main technical reasons for distinguishing between atoms and variables.

The following theorem is proved in~\cite{Brun:04:Deep-Inf:rq}, and follows immediately from Section~\ref{SectCosGentzen}, where we prove that CoS systems p-simulate Gentzen systems.

\begin{theorem}
\mbox{\rm(Br\"unnler)}\quad
Systems\/ $\SKSg$, $\KSg$, $\SKS$, and\/ $\KS$ are complete; systems\/ $\SKSg$ and\/ $\SKS$ are implicationally complete.
\end{theorem}

The theorem holds also when restricting the language to ground derivations, since systems $\SKS$ and $\KS$ apply to them.

In the presence of cut, the coweakening and cocontraction rules do not play a major role in terms of proof complexity:

\begin{theorem}
Systems\/ $\SKSg$ and\/ $\KSg\cup\giu$ are p-equivalent, and systems\/ $\SKS$ and\/ $\KS\cup\aiu$ are p-equivalent.
\end{theorem}

\begin{proof}
Observe that the rules $\gwu$ and $\gcu$ can be derived in $\KSg\cup\giu$:
\[
\vlderivation                            {
\vlin{=   }{}{                \ttt }    {
\vlin{\giu}{}{\vls[\fff      .\ttt]}   {
\vlin{\gwd}{}{\vls[(A.\bar A).\ttt]}  {
\vlin{\swi}{}{\vls[(A.\fff  ).\ttt]} {
\vlin{=   }{}{\vls(A.[\fff.\ttt])  }{
\vlhy        {     A               }}}}}}}
\qquad\hbox{and}\qquad
\vlderivation                                       {
\vlin{=   }{}{\vls                     (A.A) }     {
\vlin{\giu}{}{\vls[\fff               .(A.A)]}    {
\vlin{\gcd}{}{\vls[(A. \bar A        ).(A.A)]}   {
\vlin{\swi}{}{\vls[(A.[\bar A.\bar A]).(A.A)]}  {
\vlin{\gid}{}{\vls(A.[[\bar A.\bar A].(A.A)])} {
\vlin{=   }{}{\vls(A.\ttt                   )}{
\vlhy        {     A                         }}}}}}}}
\quad.
\]
Similar constructions hold in $\KS\cup\giu$ for $\awu$ and $\acu$.
\end{proof}

It turns out that all the systems mentioned in the previous theorem are p-equivalent, as a consequence of Corollary~\ref{CorAPSimG}.

Analytic systems are formally defined in Definition~\ref{DefSKS} for CoS, and \ref{DefGentzen} for Gentzen. Those definitions are specific to different systems in different formalisms, which is not necessarily satisfactory. Defining a general, syntax-independent concept of analyticity is a subject of ongoing research (see Problem~\ref{SubsAnCoSPSimCoS}). We briefly discuss now the notion of analyticity and its connections with the proof complexity of deep inference, as an introduction to our result on Statman tautologies.

A Gentzen system is said to be analytic when it does not contain the cut rule. Analytic Gentzen systems enjoy the `subformula property', \emph{i.e.}, proofs in these systems only contain subformulae of their conclusions. In fact, we might stipulate that enjoying the subformula property is a primitive notion of analyticity, which we can use to exclude the cut rule, as desired. In analytic Gentzen proofs, all formulae have lower or equal complexity than that of the conclusion, when complexity is measured, for example, as the and/or depth of a formula (\emph{i.e.}, the number of alternations of conjunction and disjunction; see Definition~\ref{DefDepth}). There is another property, of interest to us, that analytic Gentzen systems enjoy: given an inference rule and its conclusion, there are only finitely many premisses to choose from; we call such rules `finitely generating'. The cut rule in Gentzen does not possess the subformula property nor is it finitely generating.

The primitive notion of analyticity that we are currently adopting for CoS is different from the one for Gentzen. We stipulate that a rule is analytic if its premiss is a formula obtained from a formula scheme by instantiating it with subformulae of the conclusion (so, the premiss is not just a subformula of the conclusion). This means that no atom or variable can appear in the premiss of an analytic rule that does not appear in its conclusion. It is, of course, a weaker condition than asking for the subformula property of Gentzen systems, but doing so is necessary if we want to adopt deep inference and obtain linear rules. Like the subformula property does for Gentzen, this weaker notion for CoS excludes the cut rule, but also the coweakening one. However, this is not an important difference with the sequent calculus because coweakening is irrelevant for the proof complexity of a CoS system (see, for example \cite{GuglGund:07:Normalis:lr}). The reason for dealing with coweakening is that, given the potential importance of cocontraction for proof complexity, we preferred to introduce top-down-symmetric CoS systems (so, closed by duality), even if coweakening and cocontraction are not required for completeness.

So, the two notions of analyticity, for Gentzen and for CoS, are such that the only important rules that are not analytic are the respective cut rules. Note that in both cases, analytic systems are made of finitely generating rules. However, there is an important difference: in CoS, the complexity of formulae in an analytic proof can be unboundedly greater than the complexity of the conclusion. Consider, for example, the derivation
\[
\vlderivation                              {
\vlin{\gcd}{}{\vls [c.(a.b)]           }  {
\vlin{=   }{}{\vls[[c.(a.b)].[c.(a.b)]]} {
\vlin{\swi}{}{\vls[c.[(a.b).[c.(a.b)]]]}{
\vlhy        {\vls[c.(a.[b.[c.(a.b)])]]}}}}}
\quad.
\]
The and/or depth of the conclusion is $1$, while that of the premiss is $3$. We could repeat the construction on top of itself and further increase the and/or depth of the premiss at will.

Deep-inference systems can be top-down symmetric in the sense that a derivation can be flipped upside-down and negated and still be a valid derivation (we say that two such derivations are dual). Accordingly, some forms of analyticity can be defined in a symmetric way. Then, typically asymmetric theorems that depend on the notion of analyticity, like cut elimination, can be generalised to symmetric statements that imply cut elimination. This is not the place to be detailed about this aspect; suffice to say that we can obtain for CoS systems much stronger normalisation (and cut elimination) results than for Gentzen systems (see \cite{Brun:06:Deep-Inf:qy,GuglGund:07:Normalis:lr}).

As we see in Section~\ref{SectCosGentzen}, analyticity in CoS faithfully captures analyticity in Gentzen, in the sense that analytic CoS can produce isomorphic proofs to Gentzen ones (almost amounting to a change of notation). However, analytic CoS admits more proofs than analytic Gentzen, and, among CoS proofs, we can find some remarkably small ones, which analytic Gentzen cannot express; this is the subject of Section~\ref{SubsStat} on Statman tautologies.

\begin{remark}\label{RemLocal}
The rules of $\SKS$ are \emph{local}, in the sense that, for any language with a finite number of atoms, checking that a given expression is an instance of any of these rules requires time bounded by a constant (adopting a tree representation of formulae, for example). This property is peculiar to deep inference; it cannot be obtained in other formalisms. For example, a traditional, nonatomic contraction rule is not local because it requires checking the identity of two unbounded formulae. Contrary to other nonlocal rules, like identity in a Gentzen system, contraction cannot be replaced by its local, atomic counterpart without losing completeness. A counterexample showing this is in \cite{Brun:03:Two-Rest:mn}. Locality can possibly lead to a new, general, productive notion of analyticity, as argued in Problem~\ref{SubsAnCoSPSimCoS}.
\end{remark}

We conclude the section by showing the p-equivalence of systems with atomic rules to systems without atomic rules. We start by proving the result on ground derivations.

\begin{lemma}\label{LemGInt}
For every ground instance\/ $\vlinf{\gid}{}{\vls[\alpha.\bar\alpha]}{\ttt}$ there is a derivation\/ $\vldownsmash{\vlder{\Phi}{\{\aid,\swi\}}{\vls[\alpha.\bar\alpha]}{\ttt}}$ and for every ground instance\/ $\vldownsmash{\vlinf{\giu}{}{\fff}{\vls(\alpha.\bar\alpha)}}$ there is a derivation\/ $\vldownsmash{\vlder{\Phi}{\{\aiu,\swi\}}{\fff}{\vls(\alpha.\bar\alpha)}}$; in both cases\/ $\size{\Phi}\in\Ord{n^2}$, where $n=\size\alpha$.
\end{lemma}

\begin{proof}
Let us see the case for $\giu$, the other being its dual. We make an induction on the structure of $\alpha$. The cases when $\alpha$ is a unit or an atom are trivial: in the former case $\Phi$ consists of an instance of $=$ and in the latter the instance of $\giu$ is also an instance of $\aiu$. We only have to consider the case when $\alpha\equiv\vls[\beta.\gamma]$: we apply the induction hypothesis on the derivation
\[
\vlderivation                                                  {
\vlin{\giu}{}{\fff                                       }    {
\vlin{=   }{}{\vls(                   \gamma .\bar\gamma)}   {
\vlin{\giu}{}{\vls([\fff             .\gamma].\bar\gamma)}  {
\vlin{\swi}{}{\vls([(\bar\beta.\beta).\gamma].\bar\gamma)} {
\vlin{=   }{}{\vls((\bar\beta.[\beta.\gamma]).\bar\gamma)}{
\vlhy        {\vls([\beta.\gamma].(\bar\beta.\bar\gamma))}}}}}}}
\quad,
\]
and we obtain a derivation whose length is $\Ord{n}$, and so its size is $\Ord{n^2}$, where $n= \size\alpha$.
\end{proof}

\begin{lemma}\label{LemGWeak}
For every ground instance\/ $\vldownsmash{\vlinf\gwd{}\alpha\fff}$ there is a derivation\/ $\vldownsmash{\vlder\Phi{\{\awd,\swi\}}\alpha\fff}$ and for every ground instance\/ $\vlsmash{\vlupsmash{\vlinf\gwu{}\ttt\alpha}}$ there is a derivation\/ $\vldownsmash{\vlupsmash{\vlder\Phi{\{\awu,\swi\}}\ttt\alpha}}$; in both cases\/ $\size{\Phi}\in\Ord{n^2}$, where $n=\size\alpha$.
\end{lemma}

\begin{proof}
The proof is similar to the one for Lemma~\ref{LemGInt}. In case $\alpha\equiv\ttt$ an instance of $\gwd$ yields
\[
\vlderivation                           {
\vlin{=   }{}{                 \ttt }  {
\vlin{\swi}{}{\vls[(\fff.\fff).\ttt]} {
\vlin{=   }{}{\vls(\fff.[\fff.\ttt])}{
\vlhy        {     \fff             }}}}}
\quad;
\]
we can do similarly if $\alpha\equiv\fff$ in an instance of $\gwu$. These are the derivations for the inductive cases about $\gwd$ (those about $\gwu$ are dual):
\[
\vlderivation                       {
\vlin{\gwd}{}{\vls[\beta.\gamma]}  {
\vlin{=   }{}{\vls[\fff .\gamma]} {
\vlin{\gwd}{}{           \gamma }{
\vlhy        {           \fff   }}}}}
\qquad\hbox{and}\qquad
\vlderivation                       {
\vlin{\gwd}{}{\vls(\beta.\gamma)}  {
\vlin{\gwd}{}{\vls(\fff .\gamma)} {
\vlin{=   }{}{\vls(\fff .\fff  )}{
\vlhy        {           \fff   }}}}}
\quad.
\]
From these we obtain a derivation whose length is $\Ord{n}$, and so its size is $\Ord{n^2}$, where $n=\size\alpha$.
\end{proof}

\begin{remark}
In the statement of Lemma~\ref{LemGWeak}, instead of $\{\awd,\swi\}$ and $\{\awu,\swi\}$ we could have used $\{\awd,\med\}$ and $\{\awu,\med\}$, respectively.
\end{remark}

\begin{lemma}\label{LemGContr}
For every ground instance\/ $\vldownsmash{\vlinf\gcd{}\alpha{\vls[\alpha.\alpha]}}$ there is a derivation $\vlsmash{\vlder\Phi{\{\acd,\med\}}\alpha{\vls[\alpha.\alpha]}}$ and for every ground instance\/ $\vldownsmash{\vlinf\gcu{}{\vls(\alpha.\alpha)}\alpha}$ there is a derivation $\vldownsmash{\vlder\Phi{\{\acu,\med\}}{\vls(\alpha.\alpha)}\alpha}$; in both cases\/ $\size{\Phi}\in \Ord{n^2}$, where $n=\size\alpha$.
\end{lemma}

\begin{proof}
The proof is similar to the one for Lemma~\ref{LemGInt}. These are the derivations for the inductive cases about $\gcd$ (those about $\gcu$ are dual):
\[
\vlderivation                                        {
\vlin{\gcd}{}{\vls[ \beta       . \gamma        ]}  {
\vlin{\gcd}{}{\vls[ \beta       .[\gamma.\gamma]]} {
\vlin{=   }{}{\vls[[\beta.\beta].[\gamma.\gamma]]}{
\vlhy        {\vls[[\beta.\gamma].[\beta.\gamma]]}}}}}
\qquad\hbox{and}\qquad
\vlderivation                                        {
\vlin{\gcd}{}{\vls( \beta       . \gamma        )}  {
\vlin{\gcd}{}{\vls( \beta       .[\gamma.\gamma])} {
\vlin{\med}{}{\vls([\beta.\beta].[\gamma.\gamma])}{
\vlhy        {\vls[(\beta.\gamma).(\beta.\gamma)]}}}}}
\quad.
\]
From these we obtain a derivation whose length is $\Ord{n}$, and so its size is $\Ord{n^2}$, where $n= \size\alpha$.
\end{proof}

\begin{theorem}\label{TheoAPSimG}
For every ground\/ $\SKSg$ derivation\/ $\Phi$ there is a ground\/ $\SKS$ derivation\/ $\Phi'$ with the same premiss and conclusion of\/ $\Phi$; if $n$ is the size of\/ $\Phi$ then the size of\/ $\Phi'$ is\/ $\Ord{n^2}$; moreover, if\/ $\Phi$ is in\/ $\KSg$ then\/ $\Phi'$ is in\/ $\KS$.
\end{theorem}

\begin{proof}
The theorem follows immediately from Lemmas~\ref{LemGInt}, \ref{LemGWeak}, and \ref{LemGContr}.
\end{proof}

By Remark~\ref{RemGround}, every derivation can be `grounded', so:

\begin{corollary}\label{CorAPSimG}
$\KS$ and\/ $\KSg$ are p-equivalent and\/ $\SKS$ and\/ $\SKSg$ are p-equivalent.
\end{corollary}

\begin{remark}\label{RemMacro}
Sometimes, we use nonatomic structural rule instances in $\SKS$ and $\KS$ derivations: those instances actually stand for the $\SKS$ and $\KS$ derivations that would be obtained according to the proofs of Lemmas~\ref{LemGInt}, \ref{LemGWeak}, and \ref{LemGContr}. In this sense, we say that $\gid$, $\giu$, $\gwd$, $\gwu$, $\gcd$, and $\gcu$ are `macro' rules for $\SKS$ and $\KS$. The reason we might want to work with macro rules in $\SKS$ and $\KS$ instead of working in $\SKSg$ and $\KSg$ and then appealing to Theorem~\ref{TheoAPSimG} is to obtain finer upper bounds. This is because the size of formulae over which nonatomic structural rules operate can be much smaller than the square root of the size of a derivation, which is the pessimistic assumption of Theorem~\ref{TheoAPSimG}.
\end{remark}

\begin{remark}
All implicationally complete CoS proof systems are p-equivalent. This can be proved analogously to, or resorting to, a similar `robustness' result for Frege systems (Theorem~\ref{ThRobustness}), as argued in Remark~\ref{RemRobustness}. This means that studying proof complexity for $\SKSg$ and $\SKS$ has universal value for all CoS systems for propositional logic.
\end{remark}

\section{Calculus of\/ Structures, Gentzen Proof\/ Systems and Statman Tautologies}\label{SectStatman}

There are two parts in this section. In the first part, we show how CoS naturally p-simulates Gentzen systems, and in particular how it realizes Gentzen's notion of analyticity. In the second part, we show that analytic CoS admits polynomial proofs when analytic Gentzen only has exponential ones, in the case of Statman tautologies.

\subsection{Calculus of\/ Structures and Gentzen Proof\/ Systems}\label{SectCosGentzen}

\newcommand{\Ge }{\mathsf{Gentzen}}
\newcommand{\id }{\mathsf{id}}
\newcommand{\co }{\mathsf{c}}
\newcommand{\we }{\mathsf{w}}
\newcommand{\cut}{\mathsf{cut}}
\begin{figure}[t]
\[
\begin{array}{cccccc}
   \multispan2{$\quad\vliinf\cut {}{\phi,\psi}{\phi,A}{\bar A,\psi}$}  \\
\noalign{\smallskip}
   \multispan2{ \quad\emph{cut}}                                       \\
\noalign{\bigskip}
                     \vlinf \id  {}{A,\bar A}{}
   &            \quad\vlinf \ttt {}\ttt{}
      &         \quad\vlinf \we  {}{\phi,A}{\phi}
         &      \quad\vlinf \co  {}{\phi,A}{\phi,A,A}
            &   \quad\vlinf \vlor{}{\phi,A\vlor B}{\phi,A,B}
               &\quad\vliinf\vlan{}{\phi,A\vlan B,\psi}{\phi,A}{B,\psi}\\
\noalign{\smallskip}
                     \emph{identity}
   &            \quad\emph{true}
      &         \quad\emph{weakening}
         &      \quad\emph{contraction}
            &   \quad\emph{disjunction}
               &\quad\emph{conjunction}                                \\
\end{array}
\]
\caption{System $\Ge$.}
\label{FigGentzen}
\end{figure}

In this section, we adopt a specific one-sided (Gentzen-Sch\"utte) sequent system that we call $\Ge$ (and that is called $\mathsf{GS1p}$ in \cite{TroeSchw:96:Basic-Pr:mi}). We could have adopted any other style of presentation without affecting our results. In fact, for Gentzen systems an analogous `robustness' theorem to that for Frege systems (Theorem~\ref{ThRobustness}) can be established. This means that studying the proof complexity of $\Ge$ has universal value for the class of Gentzen systems.

\begin{definition}
Over the language of $\SKS$ formulae, the sequent-calculus proof system $\Ge$ is defined by the \emph{inference rules} in Figure~\ref{FigGentzen}, where $\phi$ and $\psi$ stand for multisets of formulae and the symbol `$,$' represents multiset union. We interpret multisets of formulae as their disjunction (where associativity is irrelevant). \emph{Derivations}, denoted by $\Delta$, are trees obtained by composing instances of inference rules; the leaves of a derivation are its \emph{premisses} and the root is its \emph{conclusion}; a derivation $\Delta$ with premisses $\phi_1$, \dots, $\phi_h$ and conclusion $\psi$ is denoted by
\[
\vltreeder{\Delta}{\psi}{\phi_1}{\!\!\!\!\dots\!\!\!\!}{\phi_h}
\quad.\]
A derivation with no premisses is a \emph{proof}. The \emph{size} $\size\Delta$ of  derivation $\Delta$ is the number of unit, atom, and variable occurrences appearing in it. In the following, every $\SKS$ formula is translated into a $\Ge$ formula in the obvious way, and vice versa; in particular, we translate a $\Ge$ multiset $\phi=\alpha_1,\dots,\alpha_h$ into $\vls[\alpha_1.\vldots.\alpha_h]$.
\end{definition}

In the language, we keep the distinction between atoms and variables because, thanks to atoms, we obtain a better upper bound for the size of Statman tautologies proofs, in the Section~\ref{SubsStat}. As we said in the case of CoS, the redundancy in the language has no consequences outside of the possibility of using certain CoS rules instead of others.

\begin{definition}\label{DefGentzen}
The proof system \emph{analytic} $\Ge$ is proof system $\Ge$ without the $\cut$ rule; \emph{analytic} derivations and proofs are those derivations and proofs in $\Ge$ where no instances of the $\cut$ rule appear.
\end{definition}

We know, of course, that both $\Ge$ and analytic $\Ge$ are complete, and that $\Ge$ proofs can be transformed into analytic $\Ge$ proofs by a cut-elimination procedure, which, in general, blows-up a given proof exponentially.

Every Gentzen derivation has natural counterparts in CoS: the idea is to (arbitrarily) sequentialise its tree structure. This is possible because the natural logical relation between tree branches is conjunction, which CoS can represent, of course. In doing so, we pay in terms of complexity because the tree structure is less redundant than CoS contexts: the size of derivations grows quadratically. Other deep-inference formalisms (currently under development, see \cite{BrunLeng:05:On-Two-F:jf,Gugl:04:Formalis:ea,Gugl:05:The-Prob:nu}) are more efficient than CoS and Gentzen formalisms in dealing with this so-called `bureaucracy'.

\begin{remark}
In the following, we assume that an empty conjunction can be represented by a nonempty conjunction of $\ttt$ units.
\end{remark}

\begin{theorem}
For every\/ $\Ge$ derivation $\Delta$ with premisses $\phi_1$, \dots, $\phi_h$ and conclusion $\psi$ there is a derivation $\vlder\Phi\SKSg\psi{\vls(\phi_1.\vldots.\phi_h)}$; if $n$ is the size of $\Delta$, the size of\/ $\Phi$ is\/ $\Ord{n^2}$; moreover, if $\Delta$ is analytic then\/ $\Phi$ is in\/ $\KSg$.
\end{theorem}

\begin{proof}
We proceed by induction on the tree structure of $\Delta$. The base cases $\vlsmash{\vlinf\id{}{A,\bar A}{}}$ and $\vlsmash{\vlinf\ttt{}\ttt{}}$ are, respectively, translated into $\vlinf\gid{}{A,\bar A}{\ttt}$ and $\ttt$. The derivations
\[
\vlderivation                             {
\vlin{\we     }{}{\phi,A               } {
\vltr{\Delta_1}  {\phi                 }{
\vlhy            {\phi_1               }}
                                        {
\vlhy            {\!\!\!\!\dots\!\!\!\!}}
                                        {
\vlhy            {\phi_h               }}}}
\quad,\qquad
\vlderivation                             {
\vlin{\co     }{}{\phi,A               } {
\vltr{\Delta_1}  {\phi,A,A             }{
\vlhy            {\phi_1               }}
                                        {
\vlhy            {\!\!\!\!\dots\!\!\!\!}}
                                        {
\vlhy            {\phi_h               }}}}
\quad,\qquad\hbox{and}\qquad
\vlderivation                             {
\vlin{\vlor   }{}{\phi,A\vlor B        } {
\vltr{\Delta_1}  {\phi,A,     B        }{
\vlhy            {\phi_1               }}
                                        {
\vlhy            {\!\!\!\!\dots\!\!\!\!}}
                                        {
\vlhy            {\phi_h               }}}}
\]
are, respectively, translated into
\[
\vlderivation                                          {
\vlin{\gwd     }{     }{\vls[\phi.A   ]            }  {
\vlin{=        }{     }{\vls[\phi.\fff]            } {
\vlde{\Phi_1   }{\SKSg}{\phi                       }{
\vlhy                  {\vls(\phi_1.\vldots.\phi_h)}}}}}
\quad,\qquad
\vlderivation                                         {
\vlin{\gcd     }{     }{\vls[\phi. A   ]           } {
\vlde{\Phi_1   }{\SKSg}{\vls[\phi.[A.A]]           }{
\vlhy                  {\vls(\phi_1.\vldots.\phi_h)}}}}
\quad,\qquad\hbox{and}\qquad
\vlderivation                                        {
\vlde{\Phi_1   }{\SKSg}{\vls[\phi.[A.B]]           }{
\vlhy                  {\vls(\phi_1.\vldots.\phi_h)}}}
\quad,
\]
where $\Phi_1$ is obtained by induction from $\Delta_1$, and some possibly necessary instances of the $=$ rule have been omitted (they depend on the exact translation of $\Ge$ multisets into $\SKSg$ formulae). The derivations
\[
{\vlnostructuressyntax
\vlderivation                              {
\vliin{\vlan   }{}{\phi,A\vlan B,\psi   } {
\vltr {\Delta_1}  {\phi,A               }{
\vlhy             {\phi_1               }}
                                         {
\vlhy             {\!\!\!\!\dots\!\!\!\!}}
                                         {
\vlhy             {\phi_h               }}}
                                          {
\vltf {\Delta_2}  {B,\psi               }{
\vlhy             {\phi_{h+1}           }}
                                         {
\vlhy             {\!\!\!\!\dots\!\!\!\!}}
                                         {
\vlhy             {\phi_k               }}
                                     {0.79}}}
}
\qquad\hbox{and}\qquad
{\vlnostructuressyntax
\vlderivation                              {
\vliin{\cut    }{}{\phi,\psi            } {
\vltr {\Delta_1}  {\phi,A               }{
\vlhy             {\phi_1               }}
                                         {
\vlhy             {\!\!\!\!\dots\!\!\!\!}}
                                         {
\vlhy             {\phi_h               }}}
                                          {
\vltf {\Delta_2}  {\bar A,\psi          }{
\vlhy             {\phi_{h+1}           }}
                                         {
\vlhy             {\!\!\!\!\dots\!\!\!\!}}
                                         {
\vlhy             {\phi_k               }}
                                     {0.79}}}
}
\]
are, respectively, translated into
\[
\vlderivation                                                       {
\vlin{\swi                     }{}{\vls[\phi.[(A.B).\psi]]    }    {
\vlin{=                        }{}{\vls[\phi.(A.[B.\psi])]    }   {
\vlin{\swi                     }{}{\vls[([B.\psi].A).\phi]    }  {
\vlin{=                        }{}{\vls([B.\psi].[A.\phi])    } {
\vlde{\vls(\Phi_1.\Phi_2)      }{}{\vls([\phi.A].[B.\psi])    }{
\vlhy                             {\vls(\phi_1.\vldots.\phi_k)}}}}}}}
\qquad\hbox{and}\qquad
\vlderivation                                                          {
\vlin{=                        }{}{\vls[\phi.            \psi ]}      {
\vlin{\giu                     }{}{\vls[\phi.[\fff      .\psi]]}     {
\vlin{\swi                     }{}{\vls[\phi.[(A.\bar A).\psi]]}    {
\vlin{=                        }{}{\vls[\phi.(A.[\bar A.\psi])]}   {
\vlin{\swi                     }{}{\vls[([\bar A.\psi].A).\phi]}  {
\vlin{=                        }{}{\vls([\bar A.\psi].[A.\phi])} {
\vlde{\vls(\Phi_1.\Phi_2)      }{}{\vls([\phi.A].[\bar A.\psi])}{
\vlhy                             {\vls(\phi_1.\vldots.\phi_k) }}}}}}}}}
\quad,
\]
where $\Phi_1$ and $\Phi_2$ are obtained by induction from $\Delta_1$ and $\Delta_2$, some possibly necessary instances of the $=$ rule have been omitted, and $\vls(\Phi_1.\Phi_2)$ stands for the derivation
\[
\vlderivation                                          {
\vlde{    \vls([\phi.A].\Phi_2)
      }{}{\vls([\phi.A].[B.\psi])                   } {
\vlde{\vls(\Phi_1.   (\phi_{h+1}.\vldots.\phi_k))
      }{}{\vls([\phi.A].(\phi_{h+1}.\vldots.\phi_k))}{
\vlhy    {\vls          (\phi_1    .\vldots.\phi_k) }}}}
\quad,
\]
where $B$ is possibly instantiated by $\bar A$; the length of this derivation and the size of the largest formula appearing in it are both $\Ord{n}$. The resulting $\Ord{n^2}$ measure of these last two cases dominates the others.
\end{proof}

\begin{corollary}
$\SKSg$ p-simulates\/ $\Ge$ and\/ $\KSg$ p-simulates analytic\/ $\Ge$.
\end{corollary}

Although it does not explicitly address complexity, \cite{Brun:04:Deep-Inf:rq} is more exhaustive than the aforesaid on the two-way translation between $\SKSg$ and $\Ge$. Translating from $\SKSg$ to $\Ge$ crucially employs the cut rule: for every inference step in $\SKSg$, a cut instance is used in $\Ge$. So, while it is very natural and easy to show that $\Ge$ p-simulates $\SKSg$ (see \cite{Brun:04:Deep-Inf:rq}), we are left with the question: does analytic $\Ge$ p-simulate $\KSg$?

\subsection{Analytic Calculus of Structures on Statman Tautologies}\label{SubsStat}

We prove here that analytic $\Ge$ does not p-simulate $\KSg$. In fact, the CoS (polynomial) inefficiency in dealing with context bureaucracy is compensated by its freedom in applying inference rules, which leads to exponential speedups on certain classes of tautologies. Here, we study Statman tautologies, which have been used to provide the classic lower bound for analytic Gentzen systems: no cut-free proofs of Statman tautologies are possible in analytic Gentzen without the proofs growing exponentially over the size of the tautologies they prove \cite{Stat:78:Bounds-f:fj}. We show that, on the contrary, $\KSg$ and $\KS$ prove Statman tautologies with polynomially growing analytic proofs.

\begin{remark}\label{RemCoContr}
The subset of $\SKSg$ only containing analytic rules is equal to $\KSg$ plus the cocontraction rule. We do not know whether cocontraction provides for exponential speedups, so separating, proof-complexity-wise, the class of $\KSg$ from that of `analytic CoS'; about this, see Problem~\ref{SubsCoContr}. In our opinion, the very notion of analyticity would benefit from some further analysis; about this, see Problem~\ref{SubsAnCoSPSimCoS}.
\end{remark}

\newcommand{\St}{\mathsf S}
\begin{definition}\label{DefStat}
For $n\ge1$, consider the following formulae:
\vlstore{%
\alpha_i  &\equiv\vls[\bar c_i.\bar d_i]     &\text{for $i\ge1$}   ,\\
\beta_k^n &\equiv\textstyle\bigwedge_{i=n}^k\alpha_i\equiv
                 \vls(\alpha_n.\beta^{n-1}_k)&\text{for $n\ge k>1$},\\
\gamma_k^n&\equiv\vls(\beta_{k+1}^n.c_k)     &\text{for $n>k\ge1$} ,\\
\delta_k^n&\equiv\vls(\beta_{k+1}^n.d_k)     &\text{for $n>k\ge1$}\vlnos.
}
\begin{align*}
\vlread
\end{align*}
Statman tautologies are, for $n\ge1$, the formulae:
\[
\St_n\equiv\vls[\bar\alpha_n.[(\gamma_{n-1}^n.\delta_{n-1}^n).[\vldots.
                                    [(\gamma_1^n.\delta_1^n).\alpha_1]\vldots]]]\quad.
\]
\end{definition}

\newcommand{\vllor}{\vlleftparbracket}
\newcommand{\vlror}{\vlrightparbracket}
\begin{example}
These are the first three Statman tautologies:
\vlstore{
\St_1\equiv\vlsbr(c_1.d_1)                    \vlor{}
          &\vlsbr    [\bar c_1.\bar d_1]      \quad,                      \\
\St_2\equiv\vlsbr(c_2.d_2)                    \vlor{}
          &\vlsbr [(([\bar c_2.\bar d_2].c_1).
                    ([\bar c_2.\bar d_2].d_1)).
                     [\bar c_1.\bar d_1]]     \quad,                      \\
\St_3\equiv\vlsbr(c_3.d_3)                    \vlor{}
          &\vllor
           \vlsbr  (([\bar c_3.\bar d_3].c_2).
                    ([\bar c_3.\bar d_3].d_2))\vlor{}                     \\
          &\vllor
           \vlsbr ((([\bar c_3.\bar d_3].[\bar c_2.\bar d_2]).c_1).
                   (([\bar c_3.\bar d_3].[\bar c_2.\bar d_2]).d_1))\vlor{}\\
          &\vlsbr                        [\bar c_1.\bar d_1]\vlror\vlror
                                                                    \quad\vlnos.
}
\begin{align*}
\vlread
\end{align*}
It is perhaps easier to understand their meaning by using implication, as in
\vlstore{
\St'_3\equiv\vlsbr[      \bar c_3.\bar d_3] \vlim{}
       &{\vlnos(}
        \vlsbr[{\vlnos(}[\bar c_3.\bar d_3] \vlim\bar c_2{\vlnos)}.
               {\vlnos(}[\bar c_3.\bar d_3] \vlim\bar d_2{\vlnos)}]\vlim{}\\
       &\vlsbr {\vlnos(}[{\vlnos(}
                       ([\bar c_3.\bar d_3].
                        [\bar c_2.\bar d_2])\vlim\bar c_1{\vlnos)}.
                         {\vlnos(}
                       ([\bar c_3.\bar d_3].
                        [\bar c_2.\bar d_2])\vlim\bar d_1{\vlnos)}]\vlim{}\\
       &\vlsbr                        [\bar c_1.\bar d_1]{\vlnos))}\quad\vlnos.
}
\begin{align*}
\vlread
\end{align*}
\end{example}

\begin{remark}\label{RemSizeStat}
$\size{\St_n}=2+2\sum_{k=1}^{n-1}\size{\gamma_k^n}+2
             =  2\sum_{k=2}^n   (\size{ \beta_k^n}+1)+4
             =2n^2+2$.
\end{remark}

It is not difficult to see why analytic $\Ge$ proofs of Statman tautologies grow exponentially (this is a classic argument that can be found in many textbooks; see, for example, \cite{ClotKran:02:Boolean-:xr}). Basically, all what analytic $\Ge$ can do while building a proof of $\St_n$ is to generate a proof tree with $\Ord{2^n}$ branches. The next lemma shows the crucial advantage of deep inference over Gentzen systems: Statman tautologies can be proved `from the inside out', which is precisely what Gentzen systems can only do by resorting to convoluted proofs involving cuts (so, nonanalytic proofs).

\begin{remark}
In the following, for brevity, we label inference steps with expressions like $n\cdot\nu$, to denote $n$ inference steps involving rule $\nu$.
\end{remark}

\begin{lemma}\label{LemStat}
For Statman tautologies\/ $\St_n$ and\/ $\St_{n+1}$ there exists a derivation $\vlsmash{\vlder{}{\KS}{\St_{n+1}}{\St_n}}$ whose length is\/ $\Ord{n}$ and size is\/ $\Ord{n^3}$.
\end{lemma}

\begin{proof}
We refer to Definition~\ref{DefStat}. The requested derivation is
\[
\Phi=\vlderivation{
\vlin{(2n-1)\cdot\gcd}{}{
\vbox{\advance\baselineskip by2pt
\hbox{$\bar\alpha_{n+1}                                          \vlor{}$\strut}
\hbox{$\vllor\vlsbr  ((\alpha_{n+1}.c_n).
                      (\alpha_{n+1}.d_n))                        \vlor{}$\strut}
\hbox{$\vllor{\vlsbr(((\alpha_{n+1}.\beta_n^n).c_{n-1}).
                     ((\alpha_{n+1}.\beta_n^n).d_{n-1}))}
                                               \vlor\vllor\vldots\vlor{}$\strut}
\hbox{$\vlsbr      [(((\alpha_{n+1}.\beta_2^n).c_1).
                     ((\alpha_{n+1}.\beta_2^n).d_1)).\alpha_1]
                                               \vldots\vlror\vlror\vlror$\strut}
}
}{\vlin{2n\cdot\swi}{}{
\vbox{\advance\baselineskip by2pt
\hbox{${\overbrace{\vlsbr[\bar\alpha_{n+1}.
                [\vldots.[\bar\alpha_{n+1}.
                          \bar\alpha_{n+1}]\vldots]]}^{2n}}      \vlor{}$\strut}
\hbox{$\vllor\vlsbr  ((\alpha_{n+1}.c_n).
                      (\alpha_{n+1}.d_n))                        \vlor{}$\strut}
\hbox{$\vllor{\vlsbr(((\alpha_{n+1}.\beta_n^n).c_{n-1}).
                     ((\alpha_{n+1}.\beta_n^n).d_{n-1}))}
                                               \vlor\vllor\vldots\vlor{}$\strut}
\hbox{$\vlsbr      [(((\alpha_{n+1}.\beta_2^n).c_1).
                     ((\alpha_{n+1}.\beta_2^n).d_1)).\alpha_1]
                                               \vldots\vlror\vlror\vlror$\strut}
}
}{\vlin{2n\cdot\gid}{}{
\vbox{\advance\baselineskip by2pt
\hbox{$\vlsbr   (([\alpha_{n+1}.\bar\alpha_{n+1}].c_n).
                 ([\alpha_{n+1}.\bar\alpha_{n+1}].d_n))         \vlor{}$\strut}
\hbox{$\vllor{\vlsbr
               ((([\alpha_{n+1}.\bar\alpha_{n+1}].\beta_n^n).c_{n-1}).
                (([\alpha_{n+1}.\bar\alpha_{n+1}].\beta_n^n).d_{n-1}))}
                                              \vlor\vllor\vldots\vlor{}$\strut}
\hbox{$\vlsbr [((([\alpha_{n+1}.\bar\alpha_{n+1}].\beta_2^n).c_1).
                (([\alpha_{n+1}.\bar\alpha_{n+1}].\beta_2^n).d_1)).
                  \alpha_1]\vldots\vlror\vlror                         $\strut}}
}{\vlhy{
\vbox{\advance\baselineskip by2pt
\hbox{$\vlsbr(c_n.d_n)                                      \vlor{}$\strut}
\hbox{$\vllor{\vlsbr((\beta_n^n.c_{n-1}).
                     (\beta_n^n.d_{n-1}))}\vlor\vllor\vldots\vlor{}$\strut}
\hbox{$\vlsbr      [((\beta_2^n.c_1).
                     (\beta_2^n.d_1)).\alpha_1]\vldots\vlror\vlror $\strut}}
}}}}}
,
\]
where we use macro inference rules as explained in Remark~\ref{TheoAPSimG}. (Example~\ref{ExCentrSt} explains the central step in the preceding derivation.) The formulae appearing in the middle of the previous derivation are the largest. Since $\size{\alpha_{n+1}}=2$ and $\size{\beta_k^n}=2(n-k+1)$, their size is $2n\cdot2+6+2(3n-3+\sum_{k=2}^n\size{\beta_k^n})+2=2n^2+8n+2$. Each $\gid$ macro inference step involves one $\swi$ and two $\aid$ steps in $\KS$, for a total of six steps, including $=$ ones; each $\gcd$ macro inference step involves one $\med$ and two $\acd$ steps in $\KS$, for a total of six steps. So, the length of $\Phi$ is $2n\cdot6+2n\cdot2+(2n-1)\cdot6=28n-6$, and so $\size\Phi\le(28n-6)(2n^2+8n+2)\in\Ord{n^3}$.
\end{proof}

Note that in the previous proof, by working with macro inference rules, we get a better upper bound for $\KS$ than if we worked in $\KSg$ and then applied Theorem~\ref{TheoAPSimG}.

\begin{theorem}\label{ThStat}
There are\/ $\KS$ proofs of Statman tautologies whose size is quadratic in the size of the tautologies they prove.
\end{theorem}

\begin{proof}
Tautology $\St_1$ is trivially provable by an instance of the $\gid$ macro rule. By repeatedly applying the previous lemma, we obtain proofs of all Statman tautologies $\St_n$, whose size is $\Ord{n^4}$. Since $\size{\St_n}\in \Ord{n^2}$ (see Remark~\ref{RemSizeStat}), the statement follows.
\end{proof}

This is enough to conclude that analytic $\Ge$ does not polynomially simulate $\KSg$ and $\KS$. Some could argue that Statman tautologies are artificial in their forcing exponential Gentzen proofs into `wildly' branching. However, notice that both notions of proof and analyticity in Gentzen systems `get into tautologies' from the outside inwards. In other words, the restricted notion of analyticity in Gentzen systems is strongly correlated to the restricted notion of proof that leads to exponential-size proofs. In CoS, both notions are more liberal, to the advantage of proof complexity. We pay a price for this in terms of proof-search complexity: there is research aimed at improving the situation, with very promising results; see \cite{Kahr:06:Nondeter:fk}.

We note that polynomial proofs on Statman tautologies are obtained by a very small dose of deep inference. In fact, the trick is done by the switch and interaction instances in the proof in Lemma~\ref{LemStat}: they all operate just below the `surface' of a formula. This leads us to state a currently open problem, in Section~\ref{SubsSpUp}.

\section{Calculus of\/ Structures and Frege Systems}\label{SectFrege}

In this section, we prove that CoS and Frege systems are p-equivalent.

In the following definitions about Frege systems, we do not assume that the language of formulae coincides with the CoS one, but, as always, this is not a very important issue.

\begin{definition}
Given a language of propositional logic formulae built over a complete base of connectives, a \emph{Frege} (\emph{proof}) \emph{system} is a finite collection of sound \emph{inference rules}, each of which is a tuple of $n>0$ formulae such that from $n-1$ \emph{premisses} one \emph{conclusion} is derived; inference rules with 0 premisses are called \emph{axioms}. Given a Frege system, a \emph{Frege derivation} of \emph{length} $l$ with \emph{premisses} $\alpha_1$, \dots, $\alpha_h$ and \emph{conclusion} $\beta_l$ is a sequence of formulae $\beta_1,\dots,\beta_l$, such that each $\beta_i$ either belongs to $\{\alpha_1,\dots,\alpha_h\}$ or is the conclusion of an instance of an inference rule whose premisses belong to $\beta_1,\dots,\beta_{i-1}$, where $1\le i\le l$; a \emph{Frege proof} of $\beta$ is a Frege derivation with no premisses and conclusion $\beta$; we use $\Upsilon$ for derivations. We require of each Frege system to be \emph{implicationally complete}, \emph{i.e.}, whenever $\vlsbr(\alpha_1.\vldots.\alpha_h)\vlim\beta$ is valid there is a derivation with premisses $\alpha_1$, \dots, $\alpha_h$ and conclusion $\beta$ in the proof system. The \emph{size} of a Frege derivation $\Upsilon$ is the number of unit, atom, and variable occurrences that it contains, and is indicated by $\size\Upsilon$.
\end{definition}

The following `robustness' theorem can easily be proved.

\begin{theorem}\label{ThRobustness}
\mbox{\rm(\emph{Robustness}, Cook-Reckhow, \cite{CookReck:79:The-Rela:mf})}\quad
All Frege systems in the same language are p-equivalent.
\end{theorem}

\newcommand{\F}{\mathsf{F}}
The theorem has been generalised by Reckhow to Frege systems in any language (under `natural translations') \cite{Reck:76:On-the-L:fk}, but we do not need this level of generality in our article. The robustness theorem allows us to work with just one Frege system, and we arbitrarily choose the following, taken from \cite{Buss:87:Polynomi:fk} and modified by adding axioms $\F_{14}$, $\F_{15}$, $\F_{16}$, and $\F_{17}$ in order to deal with units.

\newcommand  {\Fr   }{\mathsf{Frege}}
\renewcommand{\mp   }{\mathsf{mp}}
\begin{definition}
Frege system $\Fr$, over the language of formulae freely generated by units, non-negated formula variables, and the connectives $\vlor$, $\vlan$, $\vlim$, and $\vlne$, has inference rules as shown in Figure~\ref{FigFrege}, where the formulae $\F_1$, \dots, $\F_{17}$ are axioms and the inference rule $\mp$ is called \emph{modus ponens}.
\end{definition}

\newcommand  {\vllan}{\vlleftcopbracket}
\newcommand  {\vlran}{\vlrightcopbracket}
\begin{figure}[t]
\begin{multicols}{2}
Axioms:
\begin{align*}
\vls\F_1   \equiv{}&A\vlim(B\vlim\vllan A\vlan B\vlran)               \\
\vls\F_2   \equiv{}&\vllan A\vlan B\vlran\vlim A                      \\
\vls\F_3   \equiv{}&\vllan A\vlan B\vlran\vlim B                      \\
\vls\F_4   \equiv{}&A\vlim\vllor A\vlor B\vlror                       \\
\vls\F_5   \equiv{}&B\vlim\vllor A\vlor B\vlror                       \\
\vls\F_6   \equiv{}&\vlne\vlne A\vlim A                               \\
\vls\F_7   \equiv{}&A\vlim\vlne\vlne A                                \\
\vls\F_8   \equiv{}&A\vlim(B\vlim A)                                  \\
\vls\F_9   \equiv{}&\vlne A\vlim(A\vlim B)                            \\
\vls\F_{10}\equiv{}&(A\vlim(B\vlim C))\vlim
                   ((A\vlim B)\vlim(A\vlim C))\!\!\!\!\!\!\!\!\!\!\!\!\\
\vls\F_{11}\equiv{}&(A\vlim C)\vlim((B\vlim C)\vlim
                   (\vllor A\vlor B\vlror\vlim C))\!\!\!\!\!\!\!\!\!  \\
\vls\F_{12}\equiv{}&(A\vlim(B\vlim C))\vlim
                    (B\vlim(A\vlim C))                                \\
\vls\F_{13}\equiv{}&(A\vlim B)\vlim(\vlne B\vlim\vlne A)              \\
\vls\F_{14}\equiv{}&\fff\vlim\vllan A\vlan\vlne A\vlran               \\
\vls\F_{15}\equiv{}&\vllan A\vlan\vlne A\vlran\vlim\fff               \\
\vls\F_{16}\equiv{}&\ttt\vlim\vllor A\vlor\vlne A\vlror               \\
\vls\F_{17}\equiv{}&\vllor A\vlor\vlne A\vlror\vlim\ttt               \\
\end{align*}
Inference rule:
\[\vliinf{\mp}{}{B}{A}{A\vlim B}\]
\end{multicols}
\caption{System $\Fr$.}
\label{FigFrege}
\end{figure}

\begin{remark}
In the following, every $\SKS$ formula is implicitly translated into a $\Fr$ formula in the obvious way, and vice versa; in particular, we translate $\Fr$'s formulae of the kind $\alpha\vlim\beta$ into $\SKS$ formulae $\vls[\bar\alpha.\beta]$.
\end{remark}

\begin{remark}
In Frege systems, distinguishing atoms from formula variables is unnecessary because we only instantiate rules by general substitution (as opposed to renaming). So, from now on, we assume that CoS atoms correspond to Frege formula variables. Since in system $\Fr$ we have a connective for negation, we can also assume that when dual atoms and formula variables appear in an $\SKSg$ formula, their $\Fr$ translation only uses $\vlne$; for example, the $\SKSg$ formula $\vls([A.\bar A].[a.\bar a])$ is translated into $\Fr$ formula $\vls([A.\vlne A].[B.\vlne B])$ or $\vls([A.\vlne A].[\vlne B.B])$ or $\vls([\vlne A. A].[B.\vlne B])$ or $\vls([\vlne A. A].[\vlne B.B])$.\break Conversely, $\Fr$ formula $\vls([A.\vlne A].[B.\vlne B])$ is translated into $\SKSg$ or $\SKS$ formula $\vls([a.\bar a].[b.\bar b])$ or $\vls([a.\bar a].[\bar b.b])$ or $\vls([\bar a. a].[b.\bar b])$ or $\vls([\bar a. a].[\bar b.b])$ or one such formula with formula variables in the place of some of the atoms. As always, we use atoms when we need to use $\SKS$ atomic structural rules, we use formula variables when we need to instantiate formulae and derivations, and otherwise we can choose both.
\end{remark}

Translating $\Fr$ into $\SKSg$ derivations is straightforward, given that the cut rule of $\SKSg$ can easily simulate modus ponens.

\begin{theorem}\label{ThCoSPSimFrege}
For every\/ $\Fr$ derivation\/ $\Upsilon$ with premisses $\alpha_1$, \dots, $\alpha_h$, where $h\ge0$, and conclusion $\beta$, there is a derivation $\vlder\Phi\SKSg\beta{\vls(\alpha_1.\vldots.\alpha_h)}$; if $l$ and $n$ are, respectively, the length and size of\/ $\Upsilon$, then the length and size of\/ $\Phi$ are, respectively, $\Ord{l}$ and $\Ord{n^2}$.
\end{theorem}

\begin{proof}
The axioms $\F_i$ of $\Fr$ are tautologies, so each one has a proof $\Phi_i$ in $\SKSg$, for $1\le i\le17$; for example $\F_1$ and $\F_{10}$ are, respectively, proved by
\[\hss
\Phi_1=
\vlderivation                               {
\vlin{=   }{}{\vls[\bar A.[\bar B.(A.B)]]} {
\vlin{\gid}{}{\vls[[\bar A.\bar B].(A.B)]}{
\vlhy        {\ttt                       }}}}
\qquad\hbox{and}\qquad
\Phi_{10}=
\vlderivation                                                             {
\vlin{\gcd}{}{\vls[(A.(B.\bar C)).[(A.\bar B). [\bar A        .C]]]}     {
\vlin{=   }{}{\vls[(A.(B.\bar C)).[(A.\bar B).[[\bar A.\bar A].C]]]}    {
\vlin{\swi}{}{\vls[(A.(B.\bar C)).[\bar A.[[(\bar B.A).\bar A].C]]]}   {  
\vlin{\gid}{}{\vls[(A.(B.\bar C)).[\bar A.[(\bar B.[A.\bar A]).C]]]}  {  
\vlin{=   }{}{\vls[(A.(B.\bar C)).[\bar A.[(\bar B.\ttt      ).C]]]} {  
\vlin{\gid}{}{\vls[(A.(B.\bar C)).[\bar A.[ \bar B            .C]]]}{                        
\vlhy        {\ttt                                                 }}}}}}}}
\quad.
\hss\]
We proceed by induction on the length of $\Upsilon=\beta_1,\dots,\beta_k,\beta$ and we prove the existence of a derivation $\vlder{\Phi'}\SKSg{\vls((\beta_1.\vldots.\beta_k).\beta)}{\vls(\alpha_1.\vldots.\alpha_h)}$. The base case $k=0$ is as follows: 1) if $\beta$ is a premiss, then $\Phi'=\beta$; 2) if $\beta\equiv\F_i\sigma$, for some $i$ and $\sigma$, then $\Phi'=\Phi_i\sigma$. For the inductive step, given $\Upsilon_k=\beta_1,\dots,\beta_k$ and $\vlder{\Phi'_k}\SKSg{\vls(\beta_1.\vldots.\beta_k)}{\gamma_k}$, where $\gamma_k$ is the conjunction of premisses of $\Upsilon_k$, we consider the following cases:
\begin{itemize}
\item if $\beta$ is a premiss, then $\Phi'=\vlupsmash{\vlder{\vls(\Phi'_k.\beta)}{\SKSg}{\vls((\beta_1.\vldots.\beta_k).\beta)}{\vls(\gamma_k.\beta)}}$;
\item if $\beta\equiv\F_i\sigma$, for some $i$ and substitution $\sigma$, then
$\vlupsmash{\Phi'=
\vlderivation                                                         {
\vlde{\vls((\beta_1.\vldots.\beta_k).\Phi_i\sigma)
                   }{\SKSg}{\vls((\beta_1.\vldots.\beta_k).\beta )}  {
\vlin{=            }{     }{\vls((\beta_1.\vldots.\beta_k).\ttt  )} {
\vlde{\Phi'_k      }{\SKSg}{\vls (\beta_1.\vldots.\beta_k)        }{
\vlhy                      {\gamma_k                              }}}}};}$
\item if $\beta$ is the conclusion of an instance $\vldownsmash{\vliinf{\mp}{}{\beta}{\beta_{k'}}{\beta_{k'}\vlim\beta}}$, where $\beta_{k''}\equiv\beta_{k'}\vlim\beta$ and $1\le k',k''\le k$, then
\[
\Phi'=
\vlderivation                                                             {
\vlin{=   }{}{\vls((\beta_1.\vldots.\beta_k).      \beta      )}      {
\vlin{\giu}{}{\vls((\beta_1.\vldots.\beta_k).[\fff.\beta     ])}     {
\vlin{\swi}{}{\vls((\beta_1.\vldots.\beta_k).[(\beta_{k'}.
                                         \bar\beta_{k'}).\beta])}    {
\vlin{=   }{}{\vls((\beta_1.\vldots.\beta_k).(\beta_{k'}.
                                        [\bar\beta_{k'}.\beta]))}   {
\vlin{\gcu}{}{\vls(\beta_1.\vldots.
                     (\beta_{k' }.\beta_{k' }).\vldots.
                     (\beta_{k''}.\beta_{k''}).\vldots.\beta_k)}  {
\vlin{\gcu}{}{\vls(\beta_1.\vldots.
                      \beta_{k' }              .\vldots.
                     (\beta_{k''}.\beta_{k''}).\vldots.\beta_k)} {
\vlde{\Phi'_k}{\SKSg}
             {\vls(\beta_1.\vldots.\beta_{k''}.\vldots.\beta_k)}{
\vlhy        {\gamma_k                                            }}}}}}}}}
\quad,
\]
where, without loss of generality, we assumed $k'<k''$.
\end{itemize}
At every inductive step the length of the $\SKSg$ derivation is only increased by an $\Ord{1}$ number of inference steps. From $\vlder{\Phi'}{\SKSg}{\vls((\beta_1.\vldots.\beta_k).\beta)}{\vls(\alpha_1.\vldots.\alpha_h)}$ we can obtain the desired derivation $\vlupsmash{\vlder\Phi\SKSg\beta{\vls(\alpha_1.\vldots.\alpha_h)}}$ by applying once the $\gwu$ rule. So, the length of $\Phi$ is $\Ord{k}$. From this, and after inspecting the aforesaid derivations, it follows that $\size\Phi\in\Ord{k^2m}$, where $m$ is the maximum size of a formula appearing in $\Upsilon$, and so $\size\Phi\in\Ord{n^2}$, where $\size\Upsilon=n$.
\end{proof}

\begin{corollary}\label{CorSKSPSimFR}
$\SKSg$ and\/ $\SKS$ p-simulate\/ $\Fr$.
\end{corollary}

\begin{proof}
The statement for $\SKSg$ follows from Theorem~\ref{ThCoSPSimFrege}, and that for $\SKS$ from this and Corollary~\ref{CorAPSimG}.
\end{proof}

Translating derivations from $\SKSg$ to $\Fr$ requires more effort than the converse, partly because of the need to simulate deep inference, and partly because of the large `amount of inference' of $=$-rule instances. The next two lemmas take care of these two issues.

\begin{lemma}\label{LemCont}
For every\/ $\SKS$ context $\xi\vlhole$ and formulae $\alpha$ and $\beta$, there is a\/ $\Fr$ derivation with premiss $\alpha\vlim\beta$ and conclusion $\xi\{\alpha\}\vlim\xi\{\beta\}$ whose length is\/ $\Ord{m}$ and size is\/ $\Ord{n^2}$, where $m=\size{\xi\vlhole}$ and $n=\size{\xi\{\alpha\}\vlim\xi\{\beta\}}$.
\end{lemma}

\begin{proof}
Consider four $\Fr$ proofs $\Upsilon'$, $\Upsilon''$, $\Upsilon'''$, and $\Upsilon''''$, respectively of the four tautologies
\[
\vlstore{
&\vlsbr(A\vlim B)\vlim([A.C]\vlim[B.C])\quad,\qquad&
&\vlsbr(A\vlim B)\vlim([C.A]\vlim[C.B])\quad,\\
&\vlsbr(A\vlim B)\vlim((A.C)\vlim(B.C))\quad,\qquad&
&\vlsbr(A\vlim B)\vlim((C.A)\vlim(C.B))\quad\vlnos.
}
\begin{aligned}
\vlread
\end{aligned}
\]
We proceed by induction on the structure of $\xi\vlhole$. If $\xi\vlhole\equiv\xi_1\vlscn[\vlhole.\gamma_1]$, we build $\Fr$ derivation
\[
\Upsilon_1=\alpha\vlim\beta,\Upsilon'\{A\ot\alpha,B\ot\beta,C\ot\gamma_1\},
\vls[\alpha.\gamma_1]\vlim[\beta.\gamma_1]\quad;
\]
we build $\Upsilon_1$ similarly if $\xi\vlhole\equiv\xi_1\vlscn(\vlhole.\gamma_1)$ or $\xi\vlhole\equiv\xi_1\vlscn[\gamma_1.\vlhole]$ or $\xi\vlhole\equiv\xi_1\vlscn(\gamma_1.\vlhole)$. Given $\xi_1\vlhole\equiv\xi_2\vlscn[\vlhole.\gamma_2]$ or $\xi_1\vlhole\equiv\xi_2\vlscn(\vlhole.\gamma_2)$ or $\xi_1\vlhole\equiv\xi_2\vlscn[\gamma_2.\vlhole]$ or $\xi_1\vlhole\equiv\xi_2\vlscn(\gamma_2.\vlhole)$ we build $\Upsilon_2$ analogously to $\Upsilon_1$, and the premiss of $\Upsilon_2$ is the conclusion of $\Upsilon_1$. We proceed this way until we build $\Upsilon_l$, whose conclusion is $\xi\{\alpha\}\vlim\xi\{\beta\}$, where $l\le m$. We obtain the desired derivation $\Upsilon$ by concatenating $\Upsilon_1$, \dots, $\Upsilon_l$. Since the length and size of $\Upsilon'$, $\Upsilon''$, $\Upsilon'''$, and $\Upsilon''''$ are independent of $\xi\vlhole$, $\alpha$, and $\beta$, the length of $\Upsilon$ is $\Ord{m}$ and its size is $\Ord{mn}$, and so $\Ord{n^2}$.
\end{proof}

\begin{lemma}\label{LemEq}
For every\/ $\SKS$ formulae $\alpha$ and $\beta$ such that $\alpha=\beta$ there is a\/ $\Fr$ derivation with premiss $\alpha$, conclusion $\beta$, length\/ $\Ord{n^3}$, and size\/ $\Ord{n^4}$, where $n=\size\alpha+\size\beta$.
\end{lemma}

\begin{proof}
Consider the following tautologies, derived from the equations in Figure~\ref{FigEq}:
\vlstore{
\vlsbr      [A.B]&\vlsbr{}\vldi[B.A]    \quad,\qquad&
\vlsbr   [A.\fff]&\vlsbr{}\vldi A       \quad,      \\
\vlsbr      (A.B)&\vlsbr{}\vldi(B.A)    \quad,\qquad&
\vlsbr   (A.\ttt)&\vlsbr{}\vldi A       \quad,      \\
\vlsbr  [[A.B].C]&\vlsbr{}\vldi[A.[B.C]]\quad,\qquad&
\vlsbr[\ttt.\ttt]&\vlsbr{}\vldi\ttt     \quad,      \\
\vlsbr  ((A.B).C)&\vlsbr{}\vldi(A.(B.C))\quad,\qquad&
\vlsbr(\fff.\fff)&\vlsbr{}\vldi\fff     \quad,      
}
\begin{equation}\label{EqTaut}
\begin{aligned}
\vlread
\end{aligned}
\end{equation}
where each expression corresponds to the two tautologies obtained by orientating each double implication. Every such tautology can be proved in $\Fr$ with a constant-size proof, so every instance $\gamma\vlim\gamma'$ of any of these tautologies has a $\Fr$ proof of length $\Ord{1}$ and size $\Ord{m'}$, where $m'=\size\gamma+\size{\gamma'}$.  By Lemma~\ref{LemCont}, for every $\xi\vlhole$ there is a derivation with premiss $\gamma\vlim\gamma'$ and conclusion $\xi\{\gamma\}\vlim\xi\{\gamma'\}$ whose length is $\Ord{m}$ and size is $\Ord{m^2}$, where $m=\size{\xi\{\gamma\}\vlim\xi\{\gamma'\}}$. By concatenating the proof and derivation so obtained, we can build a proof of $\xi\{\gamma\}\vlim\xi\{\gamma'\}$ whose length is $\Ord{m}$ and size is $\Ord{m^2}$. By Remark~\ref{RemEq}, we can build a chain of implications
\[
\alpha\equiv\alpha_1\vlim\vldots\vlim\alpha_h\equiv\delta\equiv
\beta_k\vlim\vldots\vlim\beta_1\equiv\beta\quad,
\]
where $\delta$ is a canonical form for $\alpha$ and $\beta$, $h+k$ is $\Ord{n^2}$, and each implication $\alpha_i\vlim\alpha_{i+1}$ and $\beta_{i+1}\vlim\beta_i$ is a tautology of the form $\xi\{\gamma\}\vlim\xi\{\gamma'\}$, such that $\gamma\vlim\gamma'$ is an instance of one of the tautologies~\ref{EqTaut}. By concatenating the proofs of every $\xi\{\gamma\}\vlim\xi\{\gamma'\}$ by $\mp$, we obtain a derivation with premiss $\alpha$, conclusion $\beta$, length $\Ord{n^3}$, and size $\Ord{n^4}$.
\end{proof}

\begin{lemma}\label{LemRule}
For every inference step $\vlinf{\nu}{}\beta\alpha$, where $\nu$ is a rule of\/ $\SKSg$, there is a\/ $\Fr$ derivation with premiss $\alpha$, conclusion $\beta$, length\/ $\Ord{n}$, and size\/ $\Ord{n^2}$, where $n=\size\alpha+\size\beta$.
\end{lemma}

\begin{proof}
Each of the following tautologies, corresponding to the inference rules in Figure~\ref{FigSKSg}, can be proved in $\Fr$ with a constant-size proof:
\begin{equation}\label{RuleTaut}
\vlstore{
\vlsbr(A.\vlne A)&\vlim\fff             \quad,\qquad&
                A&\vlim\ttt             \quad,\qquad&
                A&\vlsbr\vlim(A.A)      \quad,       \\
             \fff&\vlsbr\vlim[A.\vlne A]\quad,\qquad&
             \fff&\vlim A               \quad,\qquad&
      \vlsbr[A.A]&\vlim A               \quad,       \\&&&&
  \vlsbr(A.[B.C])&\vlsbr\vlim[(A.B).C]  \quad\vlnos.
}
\begin{aligned}
\vlread
\end{aligned}
\end{equation}
Let $\vlupsmash{\vlinf{\nu}{}\beta\alpha=\vlinf{\nu}{}{\xi\{\delta\}}{\xi\{\gamma\}}}$, where $\vlupsmash{\vlinf{\nu}{}\delta\gamma}$ is an instance of $\nu$. There is a $\Fr$ proof $\Upsilon$ of $\gamma\vlim\delta$, whose length is $\Ord{1}$ and size is $\Ord{n}$, obtained by instantiating the corresponding proof to $\nu$ among those in~\ref{RuleTaut}. By Lemma~\ref{LemCont}, there exists a $\Fr$ derivation $\Upsilon'$ with premiss $\gamma\vlim\delta$, conclusion $\xi\{\gamma\}\vlim\xi\{\delta\}$, length $\Ord{n}$, and size $\Ord{n^2}$. By concatenating $\Upsilon$ and $\Upsilon'$ we obtain a proof $\Upsilon''$ of $\xi\{\gamma\}\vlim\xi\{\delta\}$. From $\Upsilon''$, by using $\mp$, we obtain the desired derivation with premiss $\alpha\equiv\xi\{\gamma\}$ and conclusion $\beta\equiv\xi\{\delta\}$.
\end{proof}

\begin{theorem}\label{ThFregePSimCoS}
For every derivation\/ $\vlder\Phi\SKSg\beta\alpha$ there is a\/ $\Fr$ derivation\/ $\Upsilon$ with premiss $\alpha$ and conclusion $\beta$; if $n$ is the size of\/ $\Phi$, then the length and size of\/ $\Upsilon$ are, respectively, $\Ord{n^4}$ and\/ $\Ord{n^5}$.
\end{theorem}

\begin{proof}
The statement immediately follows from Lemmas~\ref{LemEq} and \ref{LemRule}, after assuming that the length of $\Phi$ is $\Ord{n}$.
\end{proof}

\begin{corollary}\label{CorFRPSimSKS}
$\Fr$ p-simulates\/ $\SKSg$ and\/ $\SKS$.
\end{corollary}

\begin{remark}\label{RemRobustness}
As evidenced by the proofs of Theorems~\ref{ThCoSPSimFrege} and \ref{ThFregePSimCoS}, it does not really matter, for establishing the p-simulations, 
precisely which inference rules are adopted by the CoS and Frege systems. In fact, the simulations work because the simulating systems are implicationally complete and their set of proofs is closed under substitution. This way, the constant-size proofs in one system, simulating the rules of the other system, can be instantiated at a linear cost in order to simulate instances of rules. We can then use a robustness theorem (see Theorem~\ref{ThRobustness}) for Frege in order to establish a robustness theorem for CoS, possibly also for systems on mutually different languages: given two implicationally complete CoS systems, we p-simulate each in two appropriate Frege systems and use Frege robustness.
\end{remark}

\section{Extension and Substitution}\label{SectExt}

In this section, we show how CoS systems can be extended with the Tseitin extension rule and with the substitution rule, analogously to Frege systems. We also show the p-equivalence of all these systems, as described in the box of the diagram in the Introduction. As always, we operate under robustness theorems (relying on the mentioned one, Theorem~\ref{ThRobustness}) that ensure that the proof complexity properties we establish for the specific systems actually hold for the formalisms they belong to.

\newcommand{\xFr}{\mathsf{xFrege}}
\begin{definition}\label{DefExtFR}
An \emph{extended Frege} (\emph{proof}) \emph{system} is a Frege system augmented with the (\emph{Tseitin}) \emph{extension rule}, which is a rule with no premisses and whose instances $A\vldi\beta$ are such that the variable $A$ does not appear before in the derivation, nor appears in $\beta$ or in the conclusion of the proof. We write $A\notin\alpha$ to state that variables $A$ and $\bar A$ do not appear in formula $\alpha$. The symbol $\vldi$ stands for logical equivalence, and the specific syntax of the expressions $A\vldi\beta$ depends on the language of the Frege system in use. In the following, we consider $A\vldi\beta$ a shortcut for $\vls((A\vlim\beta).(\beta\vlim A))$. We denote by $\xFr$ the proof system where a proof is a derivation with no premisses, conclusion $\alpha_k$, and shape
\[
\alpha_1                                                   ,
\dots                                                      ,
\alpha_{i_1-1}                                             ,\,
\overbrace{A_1\vldi\beta_1}^{\mbox{$\alpha_{i_1}\equiv{}$}}\,,
\alpha_{i_1+1}                                             ,
\dots                                                      ,
\alpha_{i_h-1}                                             ,\,
\overbrace{A_h\vldi\beta_h}^{\mbox{$\alpha_{i_h}\equiv{}$}}\,,
\alpha_{i_h+1}                                             ,
\dots                                                      ,
\alpha_k                                              \quad,
\]
where all the conclusions of extension instances $\alpha_{i_1}$, \dots, $\alpha_{i_h}$ are singled out and
\[
A_1\notin\alpha_1,\dots,\alpha_{i_1-1},\beta_1,\alpha_k\quad,\qquad
\dots                                                  \quad,\qquad
A_h\notin\alpha_1,\dots,\alpha_{i_h-1},\beta_h,\alpha_k\quad,
\]
and the rest of the proof is as in $\Fr$.
\end{definition}

\begin{remark}\label{RemExtPrem}
We could have equivalently defined an $\xFr$ proof of $\alpha$ as a $\Fr$ derivation with conclusion $\alpha$ and premisses $\{A_1\vldi\beta_1,\dots,A_h\vldi\beta_h\}$ such that $A_1$, $\bar A_1$, \dots, $A_h$, $\bar A_h$ are mutually distinct and $A_1\notin\beta_1,\alpha$ and $\dots$ and $A_h\notin\beta_1,\dots,\beta_h,\alpha$. Notice that $\xFr$ is indeed a proof system in the sense that it proves tautologies. In fact, given the $\xFr$ proof just mentioned, we obtain a $\Fr$ proof by applying to it, in order, the substitutions $\sigma_h=A_h\ot\beta_h$, \dots, $\sigma_1=A_1\ot\beta_1$, and by prepending to it proofs of the tautologies $\beta_1\vldi\beta_1$, $(\beta_2\vldi\beta_2)\sigma_1$, \dots, $(\beta_h\vldi\beta_h)\sigma_{h-1}\cdots\sigma_1$. In general, a proof so obtained is exponentially bigger than the $\xFr$ one it derives from.
\end{remark}

$\SKSg$ can analogously be extended, but there is no need to create a special rule; we only need to broaden the criterion by which we recognize a proof.

\newcommand{\xSKSg}{\mathsf{xSKSg}}
\begin{definition}\label{DefExtSKSg}
An \emph{extended} $\SKSg$ \emph{proof} of $\alpha$ is an $\SKSg$ derivation with conclusion $\alpha$ and premiss $\vls([\bar A_1.\beta_1].[\bar\beta_1.A_1].\vldots.[\bar A_h.\beta_h].[\bar\beta_h.A_h])$, where $A_1$, $\bar A_1$, \dots, $A_h$, $\bar A_h$ are mutually distinct and $A_1\notin\beta_1,\alpha$ and $\dots$ and $A_h\notin\beta_1,\dots,\beta_h,\alpha$. We denote by $\xSKSg$ the proof system whose proofs are extended $\SKSg$ proofs.
\end{definition}

\begin{theorem}
For every\/ $\xFr$ proof of length $l$ and size $n$ there exists an\/ $\xSKSg$ proof of the same formula and whose length and size are, respectively, $\Ord{l}$ and\/ $\Ord{n^2}$.
\end{theorem}

\begin{proof}
Consider an $\xFr$ proof as in Definition~\ref{DefExtFR}. By Remark~\ref{RemExtPrem} and Theorem~\ref{ThCoSPSimFrege}, there exists the following $\xSKSg$ proof, whose length and size are yielded by~\ref{ThCoSPSimFrege}:
\[
\vlder{}{\SKSg}{\alpha_k}{\vls([\bar A_1.\beta_1].[\bar\beta_1.A_1].\vldots.
                               [\bar A_h.\beta_h].[\bar\beta_h.A_h])}\quad.
\]
\end{proof}

Although not strictly necessary to establish the equivalence of the four extended formalisms (see diagram in the Introduction), the following theorem is very easy to prove.

\begin{theorem}
For every\/ $\xSKSg$ proof of size $n$ there exists an\/ $\xFr$ proof of the same formula and whose length and size are, respectively, $\Ord{n^4}$ and\/ $\Ord{n^5}$.
\end{theorem}

\begin{proof}
Consider an $\xSKSg$ proof as in Definition~\ref{DefExtSKSg}. The statement is an immediate consequence of Theorem~\ref{ThFregePSimCoS}, after observing that there is an $\Ord{h}$-length and $\Ord{hn}$-size $\xFr$ proof
\[
\vls
A_1\vldi\beta_1,
\dots,
A_h\vldi\beta_h,
\dots,
((A_1\vldi\beta_1).\vldots.(A_h\vldi\beta_h))\quad.
\]
\end{proof}

\begin{corollary}\label{TheoExtFRPEqSKSX}
Systems\/ $\xFr$ and\/ $\xSKSg$ are p-equivalent.
\end{corollary}

We now move to the substitution rule.

\newcommand{\sub}{\mathsf{sub}}
\newcommand{\sFr}{\mathsf{sFrege}}
\begin{definition}
A \emph{substitution Frege} (\emph{proof}) \emph{system} is a Frege system augmented with the \emph{substitution rule} $\vlinf{\sub}{}{A\sigma}{A}$. We denote by $\sFr$ the proof system where a proof is a derivation with no premisses, conclusion $\alpha_k$, and shape
\[
\alpha_1                                                        ,
\dots                                                           ,
\alpha_{i_1-1}                                                  ,\,
\overbrace{\alpha_{j_1}\sigma_1}^{\mbox{$\alpha_{i_1}\equiv{}$}}\,,
\alpha_{i_1+1}                                                  ,
\dots                                                           ,
\alpha_{i_h-1}                                                  ,\,
\overbrace{\alpha_{j_h}\sigma_h}^{\mbox{$\alpha_{i_h}\equiv{}$}}\,,
\alpha_{i_h+1}                                                  ,
\dots                                                           ,
\alpha_k                                                   \quad,
\]
where all the conclusions of substitution instances $\alpha_{i_1}$, \dots, $\alpha_{i_h}$ are singled out, $\alpha_{j_1}\in\{\alpha_1,\dots,\alpha_{i_1-1}\}$, \dots, $\alpha_{j_h}\in\{\alpha_1,\dots,\alpha_{i_h-1}\}$, and the rest of the proof is as in $\Fr$.
\end{definition}

We rely on the following result.

\begin{theorem}
\mbox{\rm(Cook-Reckhow and Kraj\'i\v cek-Pudl\'ak, \cite{CookReck:79:The-Rela:mf,KrajPudl:89:Proposit:fk})}\quad\hfil
Systems\/ $\xFr$ and\/ $\sFr$ are p-equivalent.
\end{theorem}

We can extend $\SKSg$ with the same substitution rule as for $\Fr$. The rule is used like other proper rules of system $\SKSg$, so its instances are interleaved with $=$-rule instances.

\newcommand{\sSKSg}{\mathsf{sSKSg}}
\begin{definition}
An $\sSKSg$ proof is a proof of $\SKSg$ where, in addition to the inference steps generated by rules of $\SKSg$, we admit inference steps obtained as instances of the \emph{substitution rule} $\vlinf{\sub}{}{A\sigma}{A}$.
\end{definition}

This rule does not fit any of the usual deep-inference rule classes (see Section~\ref{SectPrel}), and (as in Frege systems) is not sound, in the sense that the premiss does not imply the conclusion. However, of course, if the premiss is provable the conclusion also is.

\begin{remark}
Notice that instances of the substitution rule cannot be used inside a context; for example, the expression on the left is not a valid $\sSKSg$ proof, while the one on the right is:
\[
\vlderivation                       {
\vlin{\sub?}{}{\vls[(B.C).\bar A]} {
\vlin{\gid }{}{\vls[A    .\bar A]}{
\vlhy         {\ttt              }}}}
\quad,\qquad
\vlderivation                               {
\vlin{\sub}{}{\vls[(B.C).[\bar B.\bar C]]} {
\vlin{\gid}{}{\vls[A    . \bar A]        }{
\vlhy        {\ttt                       }}}}
\quad.
\]
\end{remark}

In the so-called `Formalism B' of deep inference, which is currently under development \cite{Gugl:04:Formalis:ea}, and for which all the proof-complexity results in this article apply unchanged, substitution becomes part of the composition mechanism of proofs, rather than an odd extension to the set of rules.

For the time being, we can establish the promised p-equivalence of all extended systems by completing the diagram in the Introduction with the last two missing steps.

\begin{theorem}
For every\/ $\xSKSg$ proof of size $n$ there exists an\/ $\sSKSg$ proof of the same formula and whose length and size are, respectively, $\Ord{n}$ and\/ $\Ord{n^2}$.
\end{theorem}

\begin{proof}
Consider the $\xSKSg$ proof
\[
\vlder{\Phi}{\SKSg}{\alpha}{
\vls([\bar A_1.\beta_1].[\bar\beta_1.A_1].\vldots.
     [\bar A_h.\beta_h].[\bar\beta_h.A_h])}
,\qquad\!\!\!\textrm{where}\!\!\!\qquad
\begin{array}{l}
A_1\notin\beta_1,              \alpha\quad,\\
\dots                                \quad,\\
A_h\notin\beta_1,\dots,\beta_h,\alpha\quad,\!\!\!\\
\end{array}
\]
and let us call its premiss $\gamma$. We can build the $\sSKSg$ proof
\[
\vlderivation                                                      {
\vlin{=   }{     }{                             \alpha }          {
\vlin{\giu}{     }{\vls[ \fff                  .\alpha]}         {
\vlin{\gcd}{     }{\vls[ (\beta_1.\bar\beta_1) .\alpha]}        {
\vlin{\sub}{     }{\vls[[(\beta_1.\bar\beta_1).
                         (\beta_1.\bar\beta_1)].\alpha]}       {
\vlin{=   }{     }{\vls[[(A_1.    \bar\beta_1).
                         (\beta_1.\bar A_1   )].\alpha]}      {
\vlin{=   }{     }{\vdots                              }     {
\vlin{\giu}{     }{\vls[[(A_1.    \bar\beta_1).
                         (\beta_1.\bar A_1   ).\vldots.
                         \fff                 ].\alpha]}    {
\vlin{\gcd}{     }{\vls[[(A_1.    \bar\beta_1).
                         (\beta_1.\bar A_1   ).\vldots.
                         (\beta_h.\bar\beta_h)].\alpha]}   {
\vlin{\sub}{     }{\vls[[(A_1.    \bar\beta_1).
                         (\beta_1.\bar A_1   ).\vldots.
                         (\beta_h.\bar\beta_h).
                         (\beta_h.\bar\beta_h)].\alpha]}  {
\vlde{\vls[\bar\gamma.\Phi]}
           {\SKSg}{\vls[[(A_1.    \bar\beta_1).
                         (\beta_1.\bar A_1   ).\vldots.
                         (A_h.    \bar\beta_h).
                         (\beta_h.\bar A_h   )].\alpha]} {
\vlin{\gid}{     }{\vls[\bar\gamma.\gamma]             }{
\vlhy             {\ttt                                }}}}}}}}}}}}}
.
\]
\end{proof}

\begin{corollary}\label{CorSKSSubPSimExtSKS}
$\sSKSg$ p-simulates\/ $\xSKSg$.
\end{corollary}

\begin{theorem}
For every\/ $\sSKSg$ proof of size $n$ there exists a proof of the same formula in\/ $\sFr$, whose length and size are, respectively, $\Ord{n^4}$ and\/ $\Ord{n^5}$.
\end{theorem}

\begin{proof}
Every $\sSKSg$ proof has shape
\[
\vlderivation                                     {
\vlde{\Phi_h      }{\SKSg}{\alpha_{h+1}    }     {
\vlin{\sub        }{     }{\alpha_h\sigma_h}    {
\vlde{\Phi_{h-1}  }{\SKSg}{\alpha_h        }   {
\vlde{\Phi_1      }{\SKSg}{\vdots          }  {
\vlin{\sub        }{     }{\alpha_1\sigma_1} {
\vlde{\Phi_0      }{\SKSg}{\alpha_1        }{
\vlhy                     {\ttt            }}}}}}}}
\quad.
\]
By Theorem~\ref{ThFregePSimCoS}, for each of $\Phi_0$, \dots, $\Phi_h$ there exist $\Fr$ derivations $\Upsilon_0$, \dots, $\Upsilon_h$ with the same premiss and conclusion, respectively. We can then build the proof
\[
\overbrace{\dots,\alpha_1}^{\Upsilon_0}\,,\,
\overbrace{\alpha_1\sigma_1,\dots}^{\Upsilon_1}\,,
\dots,\,
\overbrace{\dots,\alpha_h}^{\Upsilon_{h-1}}\,,\,
\overbrace{\alpha_h\sigma_h,\dots,\alpha_{h+1}}^{\Upsilon_h}
\]
in $\sFr$; the cited theorem also yields its length and size.
\end{proof}

\begin{corollary}\label{CorFRSubPSimSKSSub}
$\sFr$ p-simulates\/ $\sSKSg$.
\end{corollary}

Nothing prevents us from using Tseitin extension and the substitution rule with system $\SKS$, or any other atomic or nonatomic CoS system. The integration of these mechanisms into CoS is similar to their integration into Frege systems, as the simplicity of the arguments showing p-equivalence testifies.

\section{Open Problems}\label{SectOpen}

We conclude the article with a list of open problems, some of which are currently investigated by us and other researchers.

\subsection{Relation with Resolution and Other Formalisms}

In this article, we explored the relation between CoS and Frege systems, and in the cited literature the relation between CoS and Gentzen systems has been explored in depth. There are, of course, other formalisms, like resolution, whose relation with CoS might lead to some interesting research directions. For example, the note \cite{Gugl:03:Resoluti:ce} shows how simply, compared to Gentzen systems, $\KS$ expresses resolution (analytically, of course).

\subsection{Does Cocontraction Provide for an Exponential Speedup?}\label{SubsCoContr}

As we argued in Remark~\ref{RemCoContr}, we do not know whether $\KSg$ p-simulates $\KSg\cup\{\gcu\}$, or, equivalently, whether $\KS$ p-simulates $\KS\cup\{\acu\}$.

Our intuition, as well as some clues, like the mutual behaviour of the `atomic flows' of contraction and cocontraction (see \cite{GuglGund:07:Normalis:lr}) would lead us to believe that cocontraction indeed provides for an exponential speedup. However, we know that in similar situations, like for dag-like versus tree-like Frege systems, intuition was fallacious.

If cocontraction yields an exponential speedup, we obtain an even stronger analytic system than $\KSg$, which is, in turn, stronger than analytic Gentzen. This would draw interest to a hierarchy of analytic proof systems of different strength.

Unless we prove the p-equivalence of $\KSg$ and $\KSg\cup\{\gcu\}$, we tend to consider cocontraction a simple rule-based mechanism for compressing proofs, like cut, extension, and substitution.

\subsection{Pigeonhole in Analytic CoS}

Does the pigeonhole principle, in particular in its relational variety, admit polynomially growing proofs in $\KS$? If not, does it in $\KS\cup\{\acu\}$?

Investigating this problem could be relevant to the more general, following one (Problem~\ref{SubsAnCoSPSimCoS}), about the ability of analytic CoS to simulate CoS, and so Frege. In fact, the pigeonhole principle generates some of the hardest classes of tautologies known.

We note that in \cite{Japa:07:Cirquent:fk}, Japaridze shows polynomially growing proofs for the pigeonhole class of tautologies in a deep-inference system over certain circuit-like sequents, called `cirquents'. In this case, the speedup is obtained by the sharing of logical expressions in circuits.

In \cite{Jera::On-the-C:kx}, Je\v{r}\'abek shows that there are polynomial-time constructible proofs in $\KS\cup\{\acu\}$ of the functional and onto variants of the pigeonhole principle.

\subsection{Relative Strength of Analytic CoS and CoS}\label{SubsAnCoSPSimCoS}

Some recent major progress has been made in \cite{Jera::On-the-C:kx}. There, Je\v{r}\'abek uses a construction on threshold formulae in the monotone sequent calculus, by Atserias, Galesi and Pudl\'ak \cite{AtseGalePudl:02:Monotone:yu}, to show that analytic CoS quasipolynomially simulates CoS. In \cite{BrusGuglGundPari:09:Quasipol:kx}, we provide a direct and simplified construction based on atomic flows \cite{GuglGund:07:Normalis:lr} and threshold formulae.

Because of these recent advances, we expect that analytic CoS p-simulates CoS. A more in-depth discussion of this subject is in \cite{BrusGuglGundPari:09:Quasipol:kx}. If analytic CoS p-simulates CoS, then there are polynomially growing proofs of the pigeonhole principle in analytic CoS, though not necessarily in $\KS$.

\newcommand{\faiu}{{\mathsf{fai}{\uparrow}}}
We think that investigating this problem will help us to better understand analyticity, in order to obtain for it a general and more useful definition than the one we have now. We feel that the current notion is not satisfactory because it depends on the formalism and must be defined by resorting to the syntactic structure of inference rules (or, worse, by indicating which rules are analytic and which are not).

For example, a more general, nonsyntactic definition of analyticity could be the following: a rule is analytic if, given an instance of its conclusion, the set of possible instances of the premiss is finite (this is what we call a finitely generating rule in Section~\ref{SectPrel}). In this sense, an atomic `finitary' cut rule $\vlinf{\faiu}{}{\xi\{\ttt\}}{\xi\vlscn(a.\bar a)}$, such that $a$ appears in $\xi\vlhole$, would be analytic. However, \cite{BrunGugl:04:A-First-:ys} shows that we can easily transform proofs in $\SKS$ into smaller- or equal-size proofs that only use $\faiu$ wherever $\aiu$ was used. So, we could deem $\faiu$ an analytic rule, and the system obtained from $\SKS$ by substituting $\aiu$ with $\faiu$ an analytic one, and we could immediately conclude that analytic CoS p-simulates CoS. This `solution', however, is way too cheap.

We prefer to think that $\faiu$ is not an analytic rule, in some sense to be made precise. A possible point of attack is offered by the fact that $\faiu$ is not a local rule: it requires checking that $a$ appears in its context, whose size is unbounded (see Remark~\ref{RemLocal}). So, we think it could be productive to look for a notion of analyticity that is based on boundedness instead of finiteness, and tackle the separation problem between analytic CoS and CoS under that notion. The note \cite{BrusGugl:07:On-Analy:uq} further explores this direction, but much more work is necessary.

\subsection{Strength of\/ Analytic CoS Systems Plus Substitution}

We showed that CoS and Frege systems are p-equivalent, and both remain p-equivalent when extended either with Tseitin extension or substitution. However, CoS is more flexible than Frege, because it allows to `switch off' two mechanisms that potentially provide for an exponential compression of proofs: cut and cocontraction (see Problem~\ref{SubsCoContr}).

It might be interesting to study the relative strength of systems obtained by removing from $\SKS\cup\{\sub\}$ either $\aiu$ or $\acu$ or both. (Rule $\awu$ can also be removed, but we do not see a crucial role for it.) Notice that systems $\KS\cup\{\sub\}$ and $\KS\cup\{\acu,\sub\}$ could be considered, in some sense, analytic, and we do not know their relative strength.

\subsection{Speedup of\/ Deep Inference Over Any Bounded-Depth System}\label{SubsSpUp}

We saw in Theorem~\ref{ThStat} that analytic CoS exhibits an exponential speedup over analytic Gentzen, for Statman tautologies. We argued that, in this case, the speedup is obtained by a rather trivial use of deep inference, because the depth at which inference has to be performed, in order to get the speedup, is constant. So, a natural question is whether there exists a class of tautologies that requires full-fledged deep inference in order to obtain efficient proofs. We think we found such a class, which is defined as follows.

Consider, for every propositional formula $\alpha$, the following set of second-order formulae, for $n>0$:
\vlstore{
\vlnos g(   1,\alpha)\equiv{}
  &\vlnos\forall\beta.\vlsbr[((\alpha.\beta).    \beta\strut).
                                 (\bar\beta .\bar\beta\strut)]\quad,\\
\vlnos g( n+1,\alpha)\equiv{}
  &\vlnos\forall\beta.\vlsbr[({\vlnos g(n,{\vls(\alpha.\beta)})}.
                              {\vlnos g(n,             \beta  )}\strut).
                             ({\vlnos g(n,         \bar\beta  )}.
                              {\vlnos g(n,         \bar\beta  )}\strut)]
                                                                    \quad\vlnos.
}
\begin{align*}
\vlread
\end{align*}
By using these formulae as a template, we can generate a set of first-order formulae, where the (complex) management of indices ensures their uniqueness:

\newcommand{\DT}{\mathsf{DT}}
\begin{definition}
Consider, for $m,n\ge0$
\vlstore{
\vlnos h(   1,m,\alpha)\equiv{}
  &\vls[((\alpha.\beta_{m+1}).    \beta_{m+1}\strut).
            (\bar\beta_{m+1} .\bar\beta_{m+1}\strut)]       \quad,\\
\vlnos h( n+2,m,\alpha)\equiv{}
  &\vlsbr({\vlnos h(n+1,    m,{\vls(\alpha.
                               \beta_{5^{n+1}+m})})}.
          {\vlnos h(n+1,5^n+m, \beta_{5^{n+1}+m}  )}\strut) \vlor{}\\
  &\vlsbr({\vlnos h(n+1,2\cdot 5^n+m,\bar\beta_{5^{n+1}+m})}.
          {\vlnos h(n+1,3\cdot 5^n+m,\bar\beta_{5^{n+1}+m})}\strut)\quad\vlnos.
}
\begin{align*}
\vlread
\end{align*}
Consider now
\[
f(n)\equiv h(n,0,\ttt)\quad,\qquad\mbox{for $n>0$}\quad.
\]
We define the set $\DT=\{\,f(n)\mid n>0\,\}$.
\end{definition}

The program \cite{Gugl:07:On-the-P:fk} can help in understanding the nature of these formulae.

\begin{remark}
It is not difficult to verify that $\DT$ contains tautologies possessing analytic CoS proofs that grow polynomially in the size of the tautologies.
\end{remark}

The analytic CoS proofs of the $\DT$ tautologies, when read bottom-up, work by applying interactions starting from the deepest subformulae. When this cannot be the case, we conjecture that the size of the proofs grows exponentially.

\begin{definition}\label{DefDepth}
The \emph{and/or depth} of a formula is the maximum number of alternations of conjunctions and disjunctions in the formula tree; the \emph{and/or depth} of a context $\xi\vlhole$ is the number of alternations of conjunctions and disjunctions between the hole and the root of the context tree. We define a \emph{bounded-depth} CoS proof system as a CoS proof system whose inference rules only generate inference steps at a bounded depth, namely inference steps $\vlinf\nu{}{\xi\{\delta\}}{\xi\{\gamma\}}$ are such that, if $\vlinf\nu{}\delta\gamma$ is a rule instance then the and/or depth of $\xi\vlhole$ is bounded by a given constant, and the same restriction holds for the contexts in the context closure condition of relation $=$.
\end{definition}

\begin{remark}
Note that the nonatomic rules interaction (identity), cointeraction (cut), contraction and cocontraction require establishing duality or identity of formulae of unbounded and/or depth. So, their adoption might be considered an implicit use of deep inference. However, the atomic counterparts of these rules do not suffer this problem because the `deep checking' is delegated to the inference mechanism. For this reason, proving the following conjecture is better done in the analytic part of system $\SKS$.
\end{remark}
\begin{conjecture}
In any analytic bounded-depth CoS proof system, the tautologies in $\DT$ only have proofs that grow exponentially in their size.
\end{conjecture}

\section{Conclusion}

In this article, we showed that the calculus of structures (CoS) has the same characteristics of the Frege formalism in terms of proof complexity, including when extended with Tseitin extension and substitution.

We know that, contrary to Frege, CoS has a rich proof theory, and its proof systems enjoy several properties, arguably relevant to proof complexity, that cannot be observed in other formalisms, like locality for all inference rules. We also know that other logics, like modal logics, enjoy simple and modular presentations in deep inference, which should help in proof complexity investigations. This article establishes the basic connection between proof theory in deep inference and proof complexity.

As a consequence of its flexibility in inference rule design, CoS admits a notion of analyticity that is more flexible than its counterpart for Gentzen systems. We can then explore the strength of analytic systems in finer detail than possible in Gentzen systems. In this article, we moved forward the boundary between polynomial and exponential analytic proofs by proving Statman tautologies with polynomial, analytic deep-inference proofs.

We included a list of open problems and currently active research directions.
\bigskip

\noindent\emph{Acknowledgements.} We thank Tom Gundersen and Ozan Kahramano\u gullar\i\ for having carefully read the manuscript and for having suggested several improvements.

\let\oldurl\url
\renewcommand{\url}[1]{\hfill\break\oldurl{#1}}
\bibliographystyle{alpha}
\bibliography{biblio}

\end{document}